\documentclass[reprint,amsmath, amssymb, aps, prl]{revtex4-1}
\usepackage{lipsum}  
\usepackage{stackengine}
\usepackage{graphicx}
\usepackage{amsmath}
\usepackage{soul}

\usepackage{dcolumn}
\usepackage{comment}
\usepackage{bm}
\usepackage{color}
\usepackage[normalem]{ulem}
\UseRawInputEncoding
\begin{document}
\renewcommand{\thesection}{\Roman{section}}
\preprint{APS/123-QED}

\title{Dark exciton-exciton annihilation in monolayer WSe$_2$}
\author{Daniel Erkensten$^1$, Samuel Brem$^2$, Koloman Wagner$^{3,7}$, Roland Gillen$^4$, Ra\"ul Perea-Caus\'in$^1$, Jonas D. Ziegler$^{3,7}$, Takashi Taniguchi$^5$, Kenji Watanabe$^6$, Janina Maultzsch$^4$, Alexey Chernikov$^{3,7}$, and Ermin Malic$^{2,1}$}  
\affiliation{$^1$Department of Physics, Chalmers University of Technology, 41296 Gothenburg, Sweden}
\affiliation{$^2$Department of Physics, Philipps-Universit{\"a}t Marburg, 35037 Marburg, Germany}
\affiliation{$^3$Department of Physics, University of Regensburg, D-93040 Regensburg, Germany}
\affiliation{$^4$Department of Physics, Friedrich-Alexander-Universit{\"a}t Erlangen-N\"urnberg, 91058 Erlangen-N\"urnberg, Germany}
\affiliation{$^5$International Center for Materials Nanoarchitectonics,  National Institute for Materials Science, 1-1 Namiki, Tsukuba 305-004, Japan}
\affiliation{$^6$Research Center for Functional Materials, National Institute for Materials Science, Tsukuba, 1-1 Namiki, Tsukuba 305-004, Japan}
\affiliation{$^7$Dresden Integrated Center for Applied Physics and Photonic Materials (IAPP) and W\"urzburg-Dresden Cluster of Excellence ct.qmat, Technische Universit{\"a}t Dresden, 01062 Dresden, Germany}

\begin{abstract}
The exceptionally strong Coulomb interaction in semiconducting transition-metal dichalcogenides (TMDs) gives rise to a rich exciton landscape consisting of bright and dark exciton states. At elevated densities, excitons can interact through exciton-exciton annihilation (EEA), an Auger-like recombination process limiting the efficiency of optoelectronic applications. Although EEA is a well-known and particularly important process in atomically thin semiconductors determining exciton lifetimes and affecting transport at elevated densities,
its microscopic origin has remained elusive. In this joint theory-experiment study combining microscopic and material-specific theory with time- and temperature-resolved photoluminescence measurements, we demonstrate the key role of dark intervalley states that are found to dominate the EEA rate in monolayer WSe$_2$. We reveal an intriguing, characteristic temperature dependence of Auger scattering in this class of materials with an excellent agreement between theory and experiment. Our study provides microscopic insights into the efficiency of technologically relevant Auger scattering channels within the remarkable exciton landscape of atomically thin semiconductors.\\
\end{abstract}
\maketitle
\begin{figure}[t!]
    \includegraphics[ width=\columnwidth]{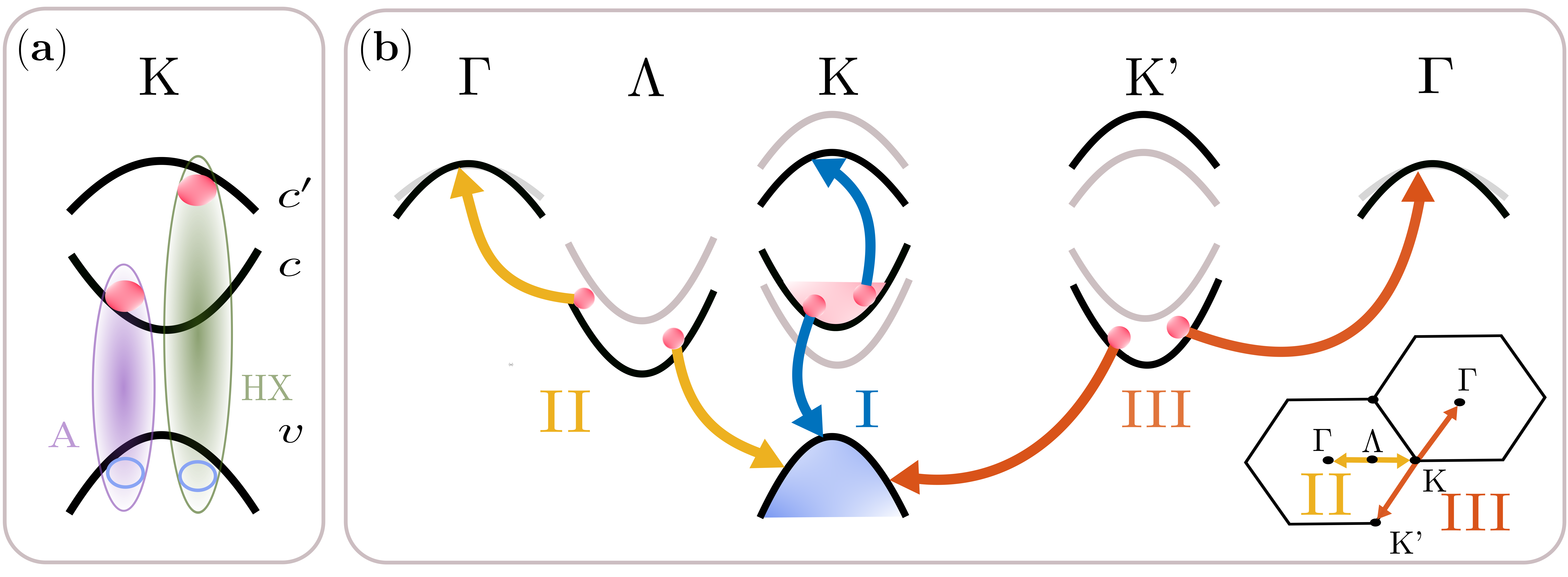}
    \caption{Schematic illustration of exciton-exciton annihilation (EEA) channels in WSe$_2$. \textbf{(a)}  The annihilation of A excitons (purple) gives rise to a higher-lying HX exciton state (green). \textbf{(b)} Regular intravalley (I blue) and additional intervalley Auger recombination processes  (II orange and III red) involving momentum-dark KK', K$\Lambda$ and K$\Gamma$ excitons, respectively.  
     The spin-split conduction bands are distinguished by black and grey lines, respectively. }
    \label{schematic}
    \end{figure}

 
Atomically thin nanomaterials, such as transition-metal dichalcogenides (TMDs),  offer an unprecedented platform to study intriguing many-particle phenomena in a broad range of external conditions \cite{wang2018colloquium,mueller2018exciton, cao2018unconventional, merkl2019ultrafast}. The weak dielectric screening and the resulting strong Coulomb interaction in these materials give rise to the formation of tightly bound excitons and promote efficient interactions between charge carriers at elevated densities. In particular, excitons can interact through exciton-exciton annihilation (EEA), an Auger recombination process shown to be very efficient in TMDs \cite{sun2014, kumar, yuan2015exciton, yu2016fundamental}. EEA is a non-radiative scattering process, in which one exciton recombines non-radiatively by transferring its energy and momentum to another exciton, resulting in a highly-excited electron-hole pair (HX) \cite{manca2017enabling, lin2020bright, lin2019quantum}, cf. Fig.\ref{schematic} (a).
The inverse process of impact excitation resulting in charge carrier multiplication has also been recently observed \cite{kim2019carrier}. Auger recombination leads to an effective saturation of exciton densities and  is thus of crucial importance for the performance of many technological applications, such as photodetectors and solar cells \cite{wang2018colloquium}.
Auger scattering has previously been shown to be extremely efficient in graphene \cite{winzer2012impact, EMCMgraphene, augernew, wendler2015impact}, but was initially considered to be inefficient in TMDs due to the difficulty to simultaneously conserve energy and momentum in parabolic band structures and the lack of resonant final states. However, recent up-converted photoluminescence (PL) measurements and ab-initio calculations have confirmed the existence of a higher energetic exciton state appearing at approximately twice the A exciton resonance both in monolayer and bilayer WSe$_2$ \cite{lin2020bright, lin2021twist}. This can be attributed to the existence of higher-lying conduction bands, enabling a particular type of resonant Auger scattering \cite{manca2017enabling, glazovprx,microscopicxx2021}, cf. Fig.\ref{schematic}. In the regular intravalley Auger recombination process discussed so far in literature, an optically excited carrier recombines with the hole at the K point and induces the excitation of another carrier into a higher conduction band (process I in Fig.\ref{schematic} (b)).

A microscopic understanding of Auger-like exciton-exciton annihilation in atomically thin semiconductors in the entire exciton landscape is still lacking. Including just the regular intravalley Auger processes turns out to be far from sufficient to explain the large EEA rates measured across different TMD materials \cite{hbnexperiment, sun2014, kumar, yuan2015exciton, yu2016fundamental}.
Phonon-assisted exciton-electron Auger recombination including dark states was also shown to be strongly suppressed \cite{Danovich_2016} and can not be responsible for the EEA efficiency seen in experiments. Furthermore, the recent observations of a strong substrate dependence of Auger scattering in TMD monolayers \cite{PhysRevB.95.241403, zipfel} still require a consistent explanation. Finally, even the bound or free nature of the highly-excited electron-hole pair that remains in the systems after EEA is not clear.
Importantly, the rich excitonic landscape of TMDs consists not only of bright intravalley excitons but also of momentum-dark intervalley excitons with non-zero center-of-mass momenta \cite{thygesendark, darkexcitons}.
These are expected to open up additional channels for Auger scattering that would satisfy momentum and energy conservation requirements. Moreover, since the energetically lowest states of tungsten-based TMDs are dark \cite{thygesendark, madeo2020directly, wallauer2021momentum,brem2020phonon}, one would expect intervalley exciton-exciton Auger recombination processes (II, III in Fig. \ref{schematic}(b)) to be particularly relevant. 

In this joint theory-experiment study, we address the nature of Auger-like exciton-exciton annihilation in atomically thin semiconductors by combining time- and temperature-resolved PL measurements with material-specific microscopic modeling including density matrix and density functional theory methods. In particular, we investigate intra- and intervalley Auger recombination channels in monolayer WSe$_2$ for different substrates and temperatures. Crucially, we show that dark intervalley Auger recombination clearly dominates the exciton-exciton annihilation. We reveal an intriguing temperature dependence, characteristic for the impact of dark states - in excellent agreement between theory and experiment. Moreover, our calculations provide insight into the previously observed decrease of Auger scattering for hBN-encapsulated WSe$_2$ and WS$_2$ monolayers\,\cite{hbnexperiment, PhysRevB.95.241403, zipfel} and we explain this effect with the changed resonance condition within the excitonic bandstructure.

\section*{Modeling of exciton-exciton annihilation rates}
To develop a realistic and material-specific approach providing microscopic insights into exciton-exciton annihilation processes in TMD monolayers, we combine first-principle calculations with the excitonic 
density matrix formalism \cite{erkensten2021exciton, brem2020phonon}. First, we  define the many-particle Hamilton operator \begin{equation}
    H_{x-x}=\frac{1}{2}\sum_{\substack{\mu\nu\rho \\ \mathbf{Q}, \mathbf{Q}'}} W^{\mu\nu\rho}_{\mathbf{Q}, \mathbf{Q}'}Y^{\dagger}_{\mu, \mathbf{Q+Q'}}X_{\nu, \mathbf{Q}}X_{\rho, \mathbf{Q}'} + \mathrm{h.c.}  
    \label{xxann}
\end{equation}
describing the annihilation of two excitons in the states $\nu$ and $\rho$ with center-of-mass momenta $\mathbf{Q}$ and $\mathbf{Q'}$ respectively, which gives rise to the formation of a single higher energetic exciton in the state $\mu$. 
The latter can be generally considered to be either a ground or an excited state with a principal quantum number $n$ that could also include unbound electron-hole pairs in the limit of $n\rightarrow \infty$.
We distinguish between excitons formed from the higher-lying conduction band $c'$ and the valence band (in the following denoted as HX excitons) and regular spin-allowed A excitons  through the exciton creation operators $X^{\dagger}=c^{\dagger}v$ and  $Y^{\dagger}=c'^{\dagger}v$, respectively, cf. Fig. \ref{schematic}(a).
Here we emphasize that HX excitons can be generally formed by electrons and holes located at different high symmetry points of the hexagonal Brillouin zone.
The compound indices $\mu, \nu$ and $\rho$ include the excitonic spin, the principal quantum numbers $n=1s,2s...$, and the excitonic valley $\xi=(\xi_h\xi_e)=\mathrm{KK^{(')}, K\Lambda, K\Gamma}$, where the first (second) letter describes the valley in which the Coulomb-bound hole (electron) is localized. In this work, we consider the hole to be at the K point, but allow the electron to be at the K, K', $\mathrm{\Lambda}$ or $\mathrm{\Gamma}$ point, cf. Fig. \ref{schematic}(b).  

\begin{figure}[t!]
    \includegraphics[width=\columnwidth]{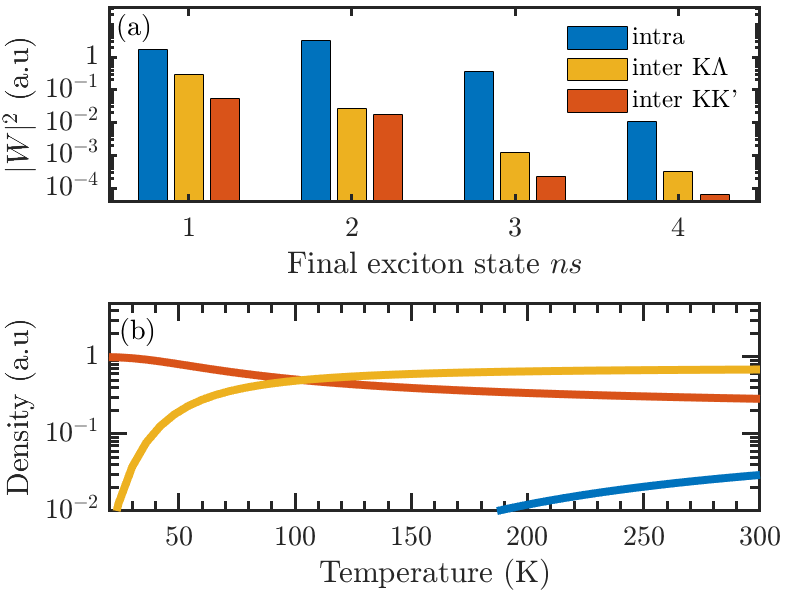}
    
    \caption{Auger matrix elements $W$ and valley-specific exciton densities in hBN-encapsulated WSe$_2$. \textbf{(a)} Intravalley and intervalley exciton Auger matrix elements $|W|^2$ evaluated at $\mathbf{Q}=\mathbf{Q}'=0$ and their dependence on the final (HX) state quantum index $n$. The matrix elements decrease rapidly with $n$, due to the small overlap between initial 1$s$ states and $n>1s$ HX states. \textbf{(b)} Temperature-dependent valley-specific exciton densities in thermal equilibrium illustrating the crucial impact of intervalley K$\Lambda$ (orange) and KK' (red) excitons and weak impact of intravalley (blue) excitons.
    }
    \label{matrixelements}
\end{figure}

The  Auger matrix element $W^{\mu\nu\rho}_{\mathbf{Q}, \bm{Q'}}$ appearing in Eq. (\ref{xxann})
determines the efficiency of the exciton-exciton annihilation process and consists of a direct and an exchange term, the latter reflecting the fermionic character of excitons. The direct and exchange components of the matrix element, which is provided in Eq. (S11) in the Supplemental Material (SM), crucially depend on the A and HX exciton wave functions (obtained from the Wannier equation \cite{haug2009quantum, kira2006many, berghauser2014analytical}) as well as the screened Coulomb interaction and electronic dipole matrix elements, with the latter being extacted from ab-initio G$_0$W$_0$ calculations. Additional details on the microscopic derivation of the Auger matrix elements and the ab-initio modeling are found in Sections IIA and III (SM) respectively.  

Fig. \ref{matrixelements}(a) illustrates the excitonic Auger matrix elements as a function of the principal quantum number $n$ of the final HX exciton state for the  intra- and intervalley processes I, II and III in  hBN-encapsulated WSe$_2$ depicted in Fig. \ref{schematic}(b). Here, we evaluated the matrix elements for $\bm{Q}=\bm{Q'}=0$, while the full momentum dependent matrix elements are provided in SM, cf. Fig. S2. We find that due to the large momentum transfer, the intervalley ($\xi_{\nu}=\xi_{\rho}=$KK', $\xi_{\mu}=$K$\Gamma$) matrix element is up to two orders of magnitude smaller than the corresponding intravalley ($\xi_{\nu}=\xi_{\rho}=\xi_{\mu}$=KK) matrix element (note the logarithmic scale). 
Note that we restrict our calculations to A excitons in the ground state $n=1s$, and vary the principal quantum number of the HX exciton in Fig. \ref{matrixelements}(a). 
We find that the matrix elements decrease rapidly with increasing quantum index $n$ due to a shrinking momentum-space overlap with the initial 1$s$ state. The only exception is the intravalley $n=2$ HX state with a slightly larger coupling than for $n=1$.
Most importantly, these results imply that the coupling to \textit{unbound} electron-hole-pairs in the limit of $n\rightarrow\infty$ should be small. 
Based on our microscopic model for the excitonic Auger matrix element, we now determine the exciton-exciton annihilation  coefficient $R_A$.

The resulting experimentally accessible recombination rates lead to an effective saturation of exciton densities and are thus important for many technological devices based on TMDs. 
We exploit the Heisenberg's equation of motion to determine the temporal evolution of the density of A excitons $n_{x}=\sum_{\nu\bm{Q}} N^{\nu}_{\mathrm{A}, \bm{Q}}$ with the momentum-dependent exciton occupation $N^{\nu}_{\mathrm{A}, \bm{Q}}=\langle X^{\dagger}_{\nu, \bm{Q}}X_{\nu, \bm{Q}}\rangle$ that we estimate by a thermal Boltzmann distribution in this work. The temperature-dependent valley-specific densities $n^{\nu}_{x}(T)=\sum_{\bm{Q}}N^{\nu}_{\mathrm{A}, \bm{Q}}$ are illustrated in Fig. \ref{matrixelements}(b), revealing that K$\mathrm{\Lambda}$ excitons dominate the density at room temperature for monolayer WSe$_2$. This reflects the energetic separation between bright and dark states (cf. Table I in SM) and the three-fold degeneracy of the $\mathrm{\Lambda}$ valley \cite{kormanyos2015k}. The latter explicitly enters the exciton distribution, crucially enhancing the occupation of K$\mathrm{\Lambda}$ excitons relative to KK and KK' excitons. 

Applying the second-order Born-Markov approximation \cite{kira2006many} (cf. Sec. IIB in SM), we find $\dot{n}_{x}=-R_A n_{x}^2$ with the exciton-exciton annihilation rate coefficient or briefly Auger coefficient $R_A$ reading 
\begin{equation}
    R_A=\frac{2\pi}{\hbar}\sum_{\substack{\mu, \nu, \rho\\\bm{Q}, \bm{Q}'}}|W^{\mu\nu\rho}_{\bm{Q}, \bm{Q}'}|^2\bar{N}^{\nu}_{\mathrm{A}, \bm{Q}}(T)\bar{N}^{\rho}_{\mathrm{A}, \bm{Q}'}(T)\delta(\Delta\epsilon) 
\label{rate}
\end{equation} 
with $\bar{N}=N/n_x$. The appearing delta function $\delta(\Delta \epsilon)$ ensures that energy is conserved during the scattering process with $\Delta \epsilon=\Delta+\epsilon^{\mu}_{\mathrm{HX}, \bm{Q+Q'}}-\epsilon^{\nu}_{\mathrm{A}, \bm{Q}}-\epsilon^{\rho}_{\mathrm{A},\bm{Q}'}$. 
The detuning $\Delta$ determines the resonance condition for the Auger scattering process that  is strongly enhanced for $\Delta\leq k_BT$.
For the intravalley Auger process ($\xi_{\mu}=\xi_{\nu}=\xi_{\rho}$=KK), the detuning  it is defined as $\Delta=E_{\mathrm{HX}}-2E_{\mathrm{A}}$, where $E_{\mathrm{HX}}$ and $E_{\mathrm{A}}$ are exciton resonance energies of the final and initial state, respectively (cf. Fig. \ref{schematic}(b)). These energies can be directly obtained from recent up-converted photoluminescence 
measurements for monolayer WSe$_2$ with $E_{\mathrm{HX}}=3.35$ eV and $E_{\mathrm{A}}=1.734$ eV \cite{lin2020bright, lin2021twist}. In the case of intervalley EEA processes, the detuning requires the knowledge of binding energies of HX and A excitons, which are microscopically calculated by solving the Wannier equation (see Sec. IIC in SM). Furthermore, we approximate the exciton dispersion as parabolic at the considered high-symmetry points, i.e. $\epsilon^{\mu}_{\bm{Q}}=\frac{\hbar^2 \bm{Q}^2}{2M^{\mu}}$ with the total exciton mass $M^{\mu}=m_e^{\mu_e}+m_h^{\mu_h}$ and the effective electron (hole) masses $m_e^{\mu_e} \ (m_h^{\mu_h})$ being extracted from first-principle calculations \cite{kormanyos2015k}. In particular, we note that the  effective mass approximation is expected to hold for the final KK HX state as shown by recent ab-initio calculations \cite{lin2020bright}.
Finally, we take into account that excitonic resonances become red-shifted with increasing temperature  \cite{varshni1967temperature, o1991temperature, arora2015excitonic}. As a result, we obtain a temperature-dependent detuning, i.e. $\Delta\rightarrow \Delta(T)=\Delta+\Delta_v(T)$ with the shift of $\Delta_v(T)=\frac{\alpha T^2}{T+\beta}$ described by the Varshni model, where the constants $\alpha$ and $\beta$ are extracted from temperature-dependent photoluminescence measurements \cite{arora2015excitonic, arora2015exciton}.  

\begin{figure*}[t!]
\includegraphics[width=0.85\textwidth]{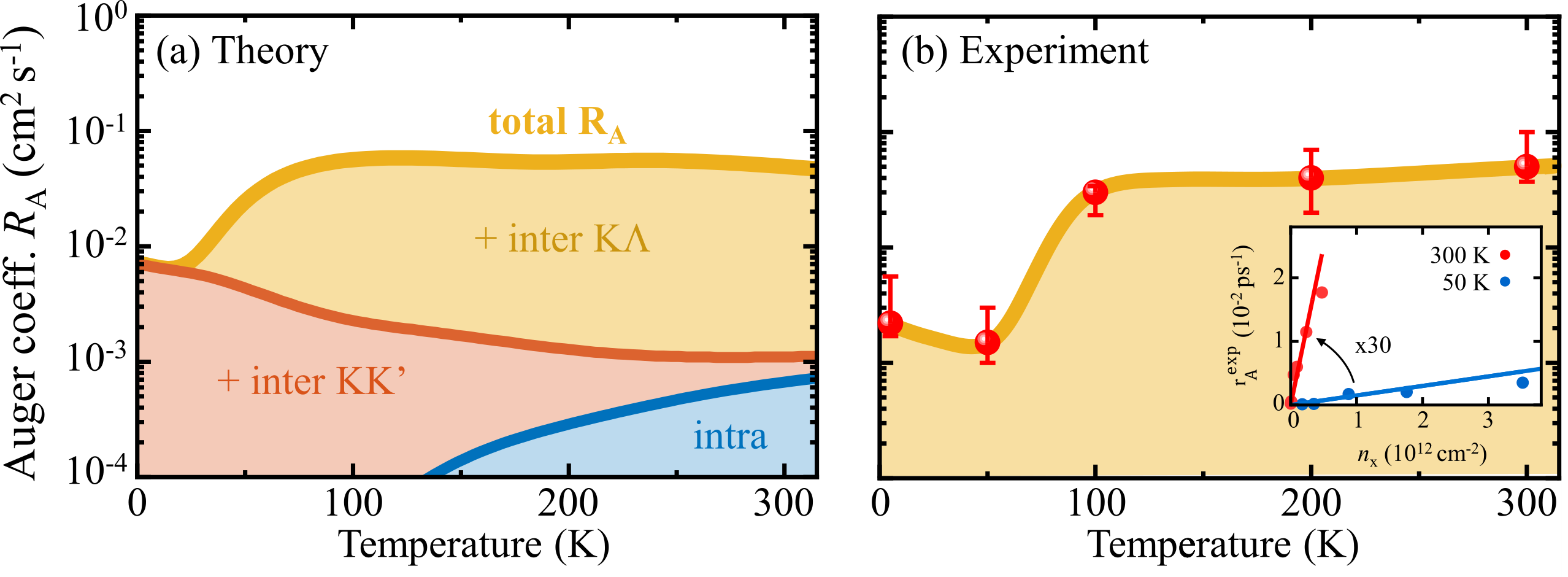}
     \caption{Temperature dependence of the Auger coefficient $R_A$ for hBN-encapsulated WSe$_2$. \textbf{(a)} Theoretically calculated $R_A$ illustrating the separate contributions of intravalley and intervalley Auger processes (KK-, KK'- and K$\mathrm{\Lambda}$, respectively), cf. Fig. \ref{schematic}(b). We reveal that the recombination of two K$\mathrm{\Lambda}$ excitons is the dominant process for temperatures above 30\,K. \textbf{(b)} Experimentally extracted Auger coefficients from time-resolved PL measurements. Error bars represent lower and upper limits for the extracted values. The inset illustrates the density-induced recombination rate $r^{\mathrm{exp}}_A$, from which the Auger coefficient can be extracted from the slope.}
    \label{tempdep}
\end{figure*}

\section*{Exciton-exciton annihilation rates}
We evaluate the exciton-exciton annihilation coefficient $R_A$ from Eq. \eqref{rate} for an hBN-encapsulated  WSe$_2$ monolayer. We take explicitly into account bright and dark A excitons  (KK, KK' and K$\mathrm{\Lambda}$) as initial states and HX excitons (KK, KK' and K$\mathrm{\Gamma}$)  up to $n=3s$ as final states for the Auger scattering process. The overlap of higher-lying states ($n>3s$) is negligibly small and thus neglected in the following, cf. Fig. \ref{matrixelements}\,(a).
In addition to matrix elements and resonance conditions, the efficiency of the Auger process is strongly determined by the distribution of excitons across lower-lying states in thermal equilibrium. Due to the energy splitting between KK, KK', and K$\mathrm{\Lambda}$ excitons being on the order of tens of meVs, the valley-specific densities strongly change with temperature, as illustrated in Fig. \ref{matrixelements}\,(b).
The resulting calculated Auger coefficients $R_A$ are presented in Fig. \ref{tempdep}\,(a) as function of temperature.

To test the theoretically predicted Auger-recombination mechanism we take advantage of density-dependent temporally resolved photoluminescence on hBN-encapsulated WSe$_2$ monolayers.
At all studied temperatures, the PL transients exhibit a density-dependent increase of the initial decay rate (cf. Fig. S1 in SM) that is well described by the bimolecular recombination law. 
Being accompanied by the saturation of the total PL intensity, this behavior is characteristic for exciton-exciton annihilation, as previously demonstrated in the literature \cite{sun2014, kumar, yuan2015exciton, yu2016fundamental}.
At each temperature, we extract the density-induced recombination rate $r_A^{\text{exp}}$ as a function of injected exciton density $n_x$ at early times after the excitation. The bimolecular coefficient attributed to Auger recombination $R_A$ is then determined from the slope, as illustrated in the inset in Fig. \ref{tempdep}\,(b) (cf. Sec. I in SM for details).
The potential contributions to the bimolecular recombination rate from biexciton formation \cite{You2015, Ye2018} can be excluded, since the latter form much faster \cite{Steinhoff2018, Nagler2018} than the observed density-dependent decay. Furthermore, we emphasize that at the considered densities the average inter-particle separation is always smaller than the diffusion length, which should be a mandatory condition for Auger scattering. In the investigated samples, long-lived dark excitons exhibit diffusion lengths on the order of 100's of nm \cite{Wagner2021}, whereas the exciton-exciton separation ranges within $30$ and $10$ nm for the studied densities between $10^{11}$ and $10^{12}$ cm$^{-2}$.

Experimentally obtained, temperature-dependent Auger coefficients are presented in Fig. \ref{tempdep}\,(b) in direct comparison to the theoretical predictions in Fig. \ref{tempdep}\,(a). 
Both, in theory and experiment the Auger coefficient increases by an order of magnitude when increasing the temperature in the range of 50 to 100 K and remains nearly constant up to room temperature. 
From microscopic calculations we obtain $R_A= 0.005\,\mathrm{cm}^2/\mathrm{s}$ and 0.05\,$\mathrm{cm}^2/\mathrm{s}$ at $T=10$ K and $T=300$ K, respectively.
These values match (without fitting) the experimentally determined coefficients of $0.001$-$0.006$\,cm$^2$/s at T $\leq 50$ K and $0.04$–$0.1$\,cm$^2$/s at $300$ K.
The obtained quantitative agreement between theory and experiment strongly supports both the predominant role of the dark excitons and the $n=1$ HX final states for Auger recombination.

To understand the microscopic origin of the drastic increase of the Auger coefficients as a function of temperature, we explicitly separate the contributions from intra- and intervalley Auger scattering in theory (corresponding to the processes I, II and III illustrated in Fig. \ref{schematic}(b)). We find that the intervalley KK' and K$\mathrm{\Lambda}$ Auger scattering involving the momentum-dark KK' and K$\Lambda$ excitons clearly dominate the Auger coefficient $R_A$ for low and high temperatures respectively (red and orange region in Fig. \ref{tempdep}(a)). The intravalley Auger scattering involving the bright KK excitons is negligible (blue region, note the logarithmic scale). This is a consequence of the spectral ordering of exciton states in WSe$_2$ monolayers, where the momentum-dark KK' excitons are the energetically lowest states followed by the dark K$\Lambda$ excitons and finally the bright KK states. The relative position of these exciton states has been determined microscopically by solving the exciton Wannier equation (see SM for more details). As a result, the dark states carry by far the largest occupation and thus dominate the Auger scattering processes, despite smaller values of the Auger matrix elements.  

The obtained temperature dependence of the Auger coefficient thus strongly reflects the changes in the valley-specific exciton densities $n_x^\nu(T)$ with $\nu=\mathrm{KK,KK',K\Lambda}$, cf.  Fig. \ref{matrixelements}\,(b). At low temperatures up to 20 K the KK' intervalley Auger process (red area in Fig. \ref{tempdep}(a))  is the predominant scattering channel. 
The K$\Lambda$ Auger scattering  takes over from 20 K and quickly becomes by far the most prominent contribution (orange area in Fig. \ref{tempdep}(a)). The sharp increase in the Auger coefficient between 50-100 K reflects the predominant population of the three-fold degenerate K$\mathrm{\Lambda}$ state at intermediate temperatures (cf. Fig. \ref{matrixelements}(a)). 
Furthermore, the Auger matrix element is much more efficient for K$\Lambda$ than KK' intervalley scattering due to the larger momentum transfer involved in the latter process suppressing the interaction (cf. Fig. \ref{matrixelements}(b).
For $T>100$ K, only minor changes in the relative exciton distributions result in approximately constant Auger coefficients at higher temperatures.

\begin{figure}[t!] 
\includegraphics[width=\columnwidth]{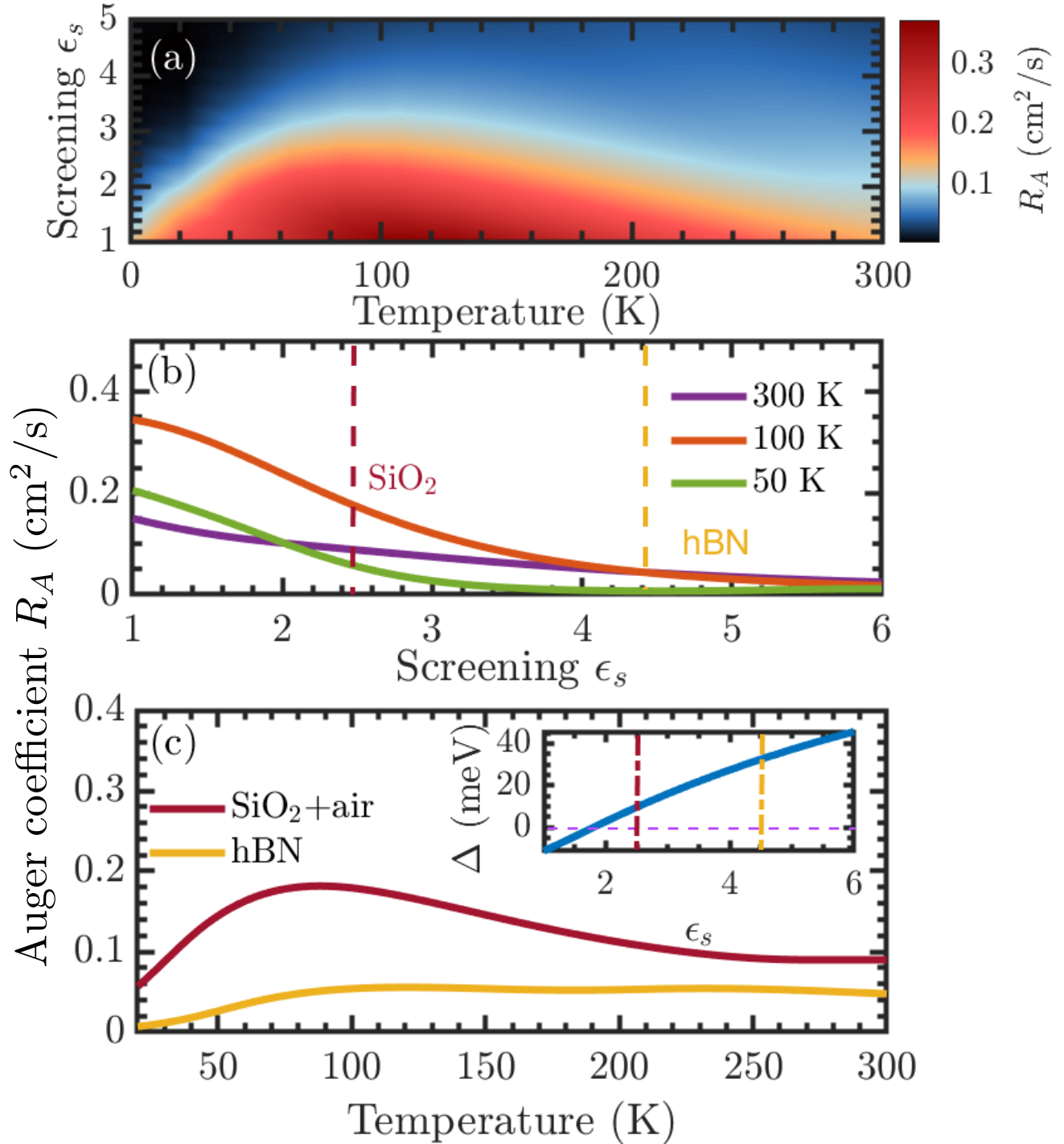}
\caption{Substrate and temperature dependence of the Auger coefficient $R_A$ in monolayer WSe$_2$ from microscopic theory. \textbf{(a)} $R_A$ as a function of screening and temperature, revealing a non-monotonic temperature behaviour. \textbf{(b)} Screening dependence of $R_A$ for different temperatures $T=50, 100$ and 300 K.  \textbf{(c)} Temperature-dependent Auger coefficients in the case of hBN-encapsulation ($\epsilon_s=4.5$) and for the SiO$_2$ substrate ($\epsilon_s=2.45$). As illustrated in the inset for the K$\mathrm{\Lambda}$ Auger process, the increase in screening from $\epsilon_s=2.45$ to $\epsilon_s=4.5$ makes the Auger scattering more off-resonant leading to a less efficient $R_A$ in hBN-encapsulated TMDs.}
\label{screentemp}
\end{figure}

Similar to the influence of temperature, the overall efficiency of the excitonic Auger scattering should also strongly depend on the dielectric environment that modifies the relative energies of the involved excitonic states.
Recently, a strong suppression of the Auger recombination in hBN-encapsulated TMD monolayers was experimentally demonstrated  \cite{PhysRevLett.120.207401, PhysRevB.95.241403}. The theoretical approach outlined above now allows us to analyze and reveal the microscopic origin of this effect. 
In the following, we investigate the impact of the dielectric environment on the exciton-exciton annihilation rate and study in particular the case of hBN-encapsulated samples versus samples placed on the standard SiO$_2$ substrate. 

In Fig.\ref{screentemp}(a), we illustrate the combined temperature and substrate dependence of the Auger coefficient $R_A$. There are two distinct trends: (i) For a fixed temperature, Auger scattering becomes less efficient with the dielectric screening, and (ii) the Auger coefficients show a maximum at a certain substrate-dependent temperature. The first trend is further shown  in Fig. \ref{screentemp}(b), where we consider the screening dependence of Auger coefficients for three different temperatures. The reduction of Auger scattering with screening is observed at all temperatures. Comparing the two most common dielectric environments, SiO$_2$ ($\epsilon_s\approx{\frac{3.9+1}{2}}=2.45$) and hBN-encapsulation ($\epsilon_s=4.5$), we find $R_{A, \mathrm{SiO}_2}=0.13$ $\mathrm{cm}^2/\mathrm{s}$ vs $R_{A, \mathrm{hBN}}=0.05$ $\mathrm{cm}^2/\mathrm{s}$ at room temperature, i.e. we predict a reduction of the Auger coefficient by approximately 60 $\%$ in the case of hBN-encapsulated WSe$_2$. This reduction can partly be attributed to a weakened Coulomb interaction with screening, but importantly it is also a consequence of  quenched resonance conditions determined by the detuning $\Delta$, cf. Eq. \eqref{rate}. To further quantify these effects, we determine the decrease in the Auger matrix element between samples on a SiO$_2$ substrate and hBN-encapsulated samples for the predominant K$\mathrm{\Lambda}$ Auger scattering channel to be approximately 30 $\%$ in hBN-encapsulated WSe$_2$ monolayers.

To understand how the resonance conditions change with the substrate we investigate the screening dependence of the individual components entering the detuning. The A resonance energy (initial state) is known to only weakly vary with dielectric screening due to the simultaneous reduction of the band gap renormalization and the excitonic binding energy \cite{bgren1, bgren2} and therefore it can be assumed to be approximately constant. Moreover, the energy splittings between different conduction bands are expected to be to a large extent independent of screening, since the Coulomb renormalization only affects the absolute energies of the bands \cite{rigid}. The detuning for the dominant K$\mathrm{\Lambda}$-Auger channel at room temperature acquires a weak screening dependence (cf. the inset of Fig. \ref{screentemp}(c)), stemming from the different screening-induced changes in intra- and intervalley binding energies. We predict $\Delta_{\mathrm{hBN}}\approx 2 \Delta_{\mathrm{SiO_2}}=30$ meV for the K$\mathrm{\Lambda}$ Auger scattering channel with the dark 1s K$\Gamma$ HX exciton as final state (cf. Fig. \ref{schematic}(b)) resulting in weaker Auger coefficients for hBN-encapsulated samples. The contribution from higher order exciton states ($n>1s$) to the Auger rates is seen to be suppressed due to large detunings (e.g. $|\Delta|_{2s, \mathrm{hBN}}\approx{200}$ meV). 
This in combination with the weakened Auger matrix elements is the origin of the previous experimental observations showing strongly quenched Auger scattering for hBN-encapsulated TMDs \cite{PhysRevLett.120.207401, PhysRevB.95.241403}.  Note that  defect-assisted Auger scattering might also play a role for TMDs on a  SiO$_2$ substrate due to disorder, while it is expected to be negligible in the case of hBN-encapsulation.  
 
Finally, we demonstrate an intriguing non-monotonic temperature dependence of Auger scattering for fixed dielectric screening in the case of SiO$_2$ and hBN-encapsulation, cf. Fig. \ref{screentemp}(c). Interestingly, we find a clear maximum in the Auger coefficient for WSe$_2$ on SiO$_2$ at around 80 K. 
The Auger coefficient $R_A$ can be approximated by $R_A\approx \frac{1}{k_B T}\mathrm{exp}(-|\Delta|/(k_B T))$, displaying a maximum at $T_{\mathrm{max}}=|\Delta|/k_B$  \cite{glazovprx}. This approximate expression becomes exact in the limit of a constant Auger matrix element and when a single exciton species $\nu=\rho\equiv \nu_0$ is dominating the Auger coefficient. 
The approximation allows us to understand the temperature dependence of the Auger coefficients, which is determined by the initial A exciton distribution and the availability of initial and final states fulfilling the conservation of energy and momentum, cf. Eq. \eqref{rate}. For very low temperatures the exciton distribution is strongly localized at vanishing kinetic energies and hence, the EEA is inefficient due to the non-zero energy detuning of initial and final states. With increasing temperature the momentum-dependent distribution becomes broadened facilitating energy and momentum conservation and thus offering additional scattering channels leading to an enhanced $R_A$. A further increase in temperature leads to a redistribution of excitons and an overall decrease in the initial population of exciton states, resulting in a reduction of the Auger coefficients. The interplay of those two effects is the origin of the observed maximum in the $R_A$ at certain intermediate temperatures.
In the case of hBN-encapsulated samples we do not observe a pronounced maximum of the Auger coefficient as a function of temperature, as the resonance energy can not be reached in the considered temperature range, i.e. $|\Delta|> k_B T$ at all temperatures. Here, the large scattering efficiency at higher temperatures is solely determined by the K$\mathrm{\Lambda}$ exciton occupation. 
\section*{Conclusions}
In this joint theory-experiment study combining microscopic, material-specific modelling with time-resolved photoluminescence measurements, we establish a fundamental understanding of Auger-like exciton-exciton annihilation processes in atomically thin semiconductors.  
We demonstrate the key importance of dark intervalley excitons in the prototypical WSe$_2$ material resulting in an intriguing temperature dependence of Auger processes. We find an excellent qualitative and quantitative agreement between theory and experiment without any adjusted free parameters. Our results also contribute to resolving an open question in the literature regarding the origin of consistently observed suppression of exciton-exciton annihilation upon hBN-encapsulation.
Overall, our work provides microscopic insights into the many-particle processes behind the technologically important exciton-exciton annihilation channels in atomically thin semiconductors. The developed approach can be further generalized to van der Waals heterostructures and twisted moir\'e exciton systems.
\section*{Acknowledgments}
We thank Maja Feierabend (Chalmers) and Paulo Eduardo de Faria Junior and Kai-Qiang Lin (University of Regensburg) for fruitful discussions. This project has received funding from  Deutsche Forschungsgemeinschaft via CRC 1083 (project B09), Emmy Noether Initiative (CH 1672/1), CRC 1277 (project B05), CRC 953 (project B13), the European Unions Horizon 2020 research and innovation programme under grant agreement no. 881603 (Graphene Flagship) and the W\"urzburg-Dresden Cluster of Excellence on Complexity and Topology in Quantum Matter ct.qmat (EXC 2147, Project-ID: 390858490). Furthermore, we are thankful to Vinnova for the support via the 2D-TECH competence center. K.Watanabe and T.T. acknowledge support from the Elemental Strategy Initiative conducted by the MEXT, Japan (Grant Number JPMXP0112101001) and  JSPS
KAKENHI (Grant Numbers JP19H05790 and JP20H00354).

\end{document}


\renewcommand{\theequation}{S\arabic{equation}}
\renewcommand{\thesection}{\Roman{section}}
\preprint{APS/123-QED}

\title{Supplemental Material for \--- Dark exciton-exciton annihilation in monolayer WSe$_2$}
\title{\textbf{Supplemental Material for} \\Dark exciton-exciton annihilation in monolayer WSe$_2$}
\author{Daniel Erkensten$^1$, Samuel Brem$^2$, Koloman Wagner$^{3,7}$, Roland Gillen$^4$, Ra\"ul Perea-Caus\'in$^1$, Jonas D. Ziegler$^{3,7}$, Takashi Taniguchi$^5$, Kenji Watanabe$^6$, Janina Maultzsch$^4$, Alexey Chernikov$^{3,7}$, and Ermin Malic$^{2,1}$}
\affiliation{$^1$Department of Physics, Chalmers University of Technology, 41296 Gothenburg, Sweden}
\affiliation{$^2$Department of Physics, Philipps-Universit{\"a}t Marburg, 35037 Marburg, Germany}
\affiliation{$^3$Department of Physics, University of Regensburg, D-93040 Regensburg, Germany}
\affiliation{$^4$Department of Physics, Friedrich-Alexander-Universit{\"a}t Erlangen-Nürnberg, 91058 Erlangen-Nürnberg, Germany}
\affiliation{$^5$International Center for Materials Nanoarchitectonics,  National Institute for Materials Science, 1-1 Namiki, Tsukuba 305-004, Japan}
\affiliation{$^6$Research Center for Functional Materials, National Institute for Materials Science, Tsukuba, 1-1 Namiki, Tsukuba 305-004, Japan}
\affiliation{$^7$Dresden Integrated Center for Applied Physics and Photonic Materials (IAPP) and Würzburg-Dresden Cluster of Excellence ct.qmat, Technische Universität Dresden, 01062 Dresden, Germany}
\maketitle 
\date{\today}

\section{Measurements of Auger coefficients} 

\begin{figure}[t!]
        \centering
        \includegraphics[width=\linewidth]{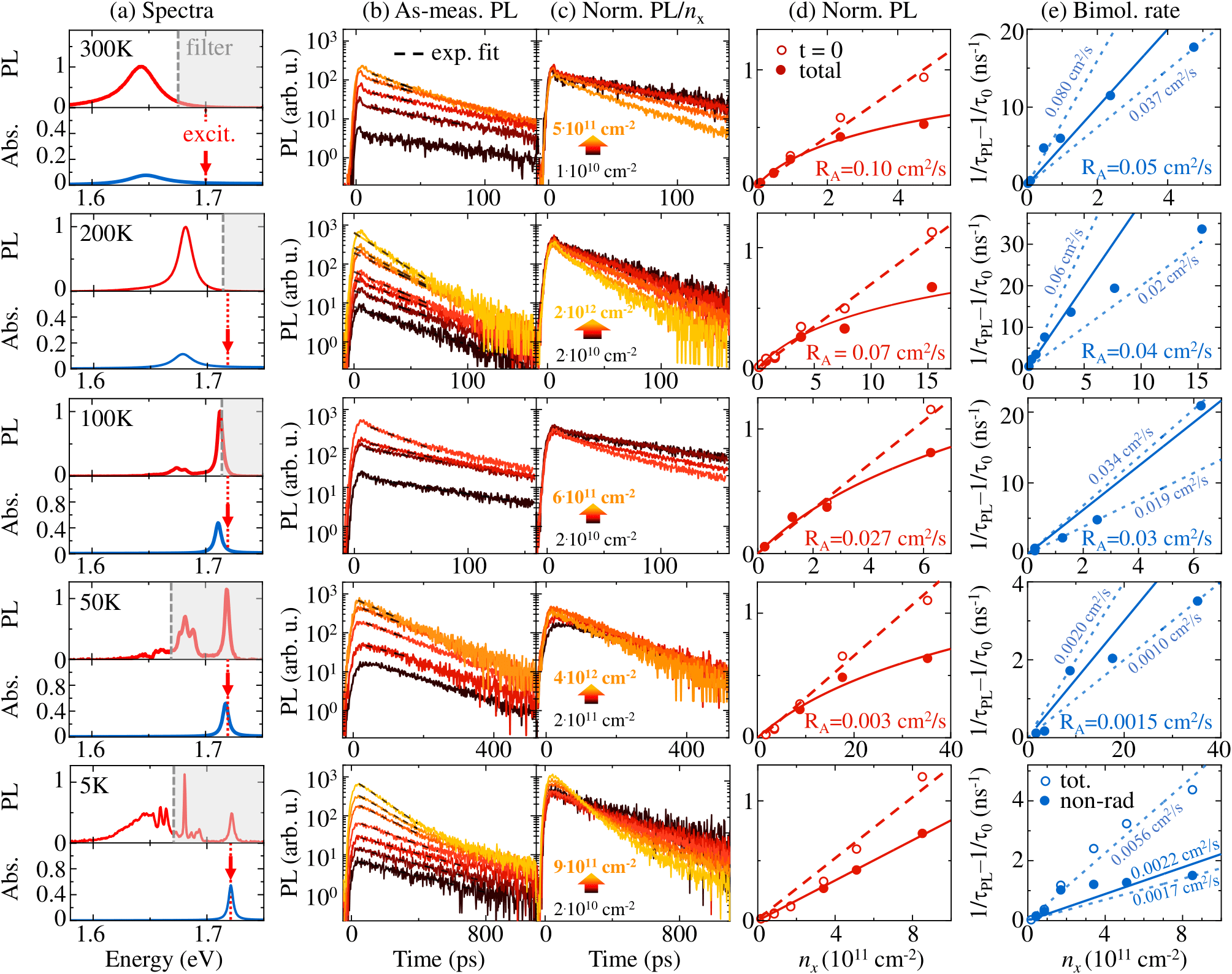}
        \caption{Analysis of temperature-dependent Auger  coefficients. Top, middle and bottom panels correspond to T $=$ $300$ K, $100$ K and $50$ K, respectively. 
        \textbf{(a)} Normalized time-integrated PL spectra after continuous-wave excitation at $2.326$\,eV (red line) and simulated absorbance from reflectance contrast measurements (blue line). Red arrows indicate excitation energy for time-resolved measurements. Grey areas mark spectrally filtered regions by applying a long pass filter. \textbf{(b)} PL transients after pulsed excitation and spectral filtering as indicated in (a). The emission is collected from the center of the PL spot. Dashed black lines indicate mono exponential fits of the initial decay.
        \textbf{(c)} Data shown in (b) normalized to electron-hole pair density $n_{eh}$ (rescaled by a constant factor).
        \textbf{(d)} Total photoluminescence $I_{PL}(total)$ and initial photoluminescence $I_{PL}(t=0)$ as function of density, extracted from (b). Solid red lines correspond to phenomenological fits with Auger coefficient R$_A$. Dashed red lines emphasize linear density dependence. 
         \textbf{(e)} Density-dependent component of the recombination rate as function of exciton density extracted from the initial photoluminescence decay. $1/\tau_{PL}$ is estimated by monoexponential fits in (b) and subtracted by the zero-density recombination constant $1/\tau_{0}$. The Auger  coefficient R$_A$ is determined by linear fits. Solid blue lines correspond to best fit, dashed lines to min. and max. fit. At T $=$ $5$ K the Auger coefficient is obtained by additionally subtracting the inferred density-induced increase of the radiative recombination rate, determined from I$_{PL}(t=0)$.}
        \label{fig2}
    \end{figure}

The Auger coefficients shown in the main manuscript are obtained from analyzing both transient luminescence as well as the total yield as function of injected exciton density.
Fig. \ref{fig2} shows a detailed overview of this analysis, including PL and absorbance spectra extracted from reflectance, emission transients and total PL intensity at selected temperatures. The individual steps of the applied measurement procedure are outlined in the following.

\subsection{Density estimation}

The sample is excited by a pulsed $140$ fs Ti:sapphire laser with repetition rate $f_{rep}=80$ MHz and photon energy E$_{ph}=1.726$ eV (resonant to the bright KK exciton at T $=$ $5$ K) focused to a spot with a diameter of FWHM = 1 $\mu$m. Effective, temperature dependent absorption is taken into account by performing spectrally-resolved reflectance contrast measurements at each studied temperature, analyzed by the transfer-matrix formalism \cite{Byrnes2016} with model dielectric functions \cite{Raja2019, Brem2019}. Accordingly, the blue curves in Fig. \ref{fig2}(a) show the extracted absorbance as function of photon energy. The average exciton density $n_{x}$ is then calculated by
\begin{equation}
n_{x}=\frac{\text{P}\alpha}{f_{rep} \text{E}_{ph} \text{A}} 
\end{equation}
with excitation power P and spot area A$=$FWHM$^2\pi/4$. 
The effective absorption $\alpha$ is determined by additionally weighting the extracted absorbance with the spectral overlap of the excitation laser. 

\subsection{Spectral filtering}

To match the theoretical scenario as close as possible we take advantage of spectrally filtering individual emissions in the experiment. At T $\leq 50$ K only phonon-assisted emissions from neutral dark excitons are considered \cite{brem2020phonon, Liu2019, He2020}. Particularly, emissions related to density-induced many body complexes, such as neutral and charged biexcitons, are excluded. At elevated temperatures weak emissions from charged excitons contribute to the signal, cf. Fig. \ref{fig2}(a). Due to the weak contribution, their impact to the overall behavior of the bimolecular recombination rate is expected to be negligible. 
Here, we note that regardless of the predominant emission channel, the majority of exciton population in WSe$_2$ monolayers should reside in lower-lying KK' and K$\Lambda$ dark states at all studied temperatures.  

\subsection{Auger coefficients extracted from total PL}

First, Auger coefficients R$_A$ are determined from the total luminescence yield I$_{PL}$. Considering bimolecular recombination processes the total intensity I$_{PL}$ is expected to saturate with increasing density. For a quantitative description we consider \cite{Kulig2018}
\begin{equation}
\text{I}_{PL}=\frac{1}{\tau_0}\frac{\text{ln}(1+n_{x}\text{R}\tau_0)}{\text{R}_A}
\label{totalPL} 
\end{equation}
with average exciton density $n_{x}$ and zero-density recombination time constant $\tau_0$.
Filled dots in Fig. \ref{fig2}(d) show the total time-integrated PL from the spot center as function of electron-hole pair density. Fitting the data with Eq. (\ref{totalPL}) and fixed $\tau_0$, corresponding to the independently determined zero-density decay time (determined in the low density regime where the decay time is constant), results in Auger  coefficients between $0.027$ and $0.1$ cm$^2/s$ for T $=$ $100 - 300$ K and an order of magnitude lower Auger coefficient of $0.003$ cm$^2/s$ at $50$ K. It should be noted that at T $=$ $5$ K the total luminescence does not show saturation and I$_{PL}(t=0)$ exhibits a super linear behavior, potentially indicating either non-linear dependence of the exciton population on excitation power or increasing radiative rates. 

\subsection{Auger coefficients extracted from emission decay}

For a more accurate determination of the temperature-dependent Auger rates we take advantage of the density-induced PL decay. Considering time-resolved measurements the density-induced emission decay rate $1/\tau_{PL}$ is estimated by mono exponential fits as indicated in Fig. \ref{fig2}(b) by black dashed lines. Fig. \ref{fig2}(e) shows the corresponding density-induced recombination rates after subtracting the zero-density rate $1/\tau_0$. The Auger coefficients R$_A$ are then extracted from linear fits to the data (blue lines). Estimated limits of R$_A$ are indicated by dashed lines. Consistent with the values determined from the total PL yield,  the Auger  coefficients are between $0.03$ and $0.05$ cm$^2/s$ at $\geq 100$ K, whereas  at T $=$ $50$ K we obtain a by more than an order of magnitude smaller Auger coefficient of $0.0015$ cm$^2/s$. 
At T $=$ $5$ K a bimolecular recombination rate of $0.0022$ cm$^2/s$ would be obtained by accounting for the inferred density-induced increase of the \textit{radiative} recombination rate, determined from I$_{PL}(t=0)$.

\section{Microscopic modeling}
\subsection{Excitonic Auger matrix elements} 
Here, we show how the excitonic Auger Hamilton operator $H_{\mathrm{x-x}}$ can be extracted from the conventional Auger Hamiltonian in the electron-hole picture.  Our strategy is to first derive the Heisenberg equation of motion for the microscopic polarization in electron-hole picture and rewrite the result in an excitonic basis, and second to assume an ansatz for a Hamiltonian in the excitonic picture and construct the corresponding excitonic matrix element such that it corresponds to the first result. In the electron-hole picture we can straightforward write down the Auger Hamilton operator corresponding to electron-mediated Auger scattering:
\begin{equation}
    H_{\mathrm{Aug}}=\sum_{\substack{\xi_1...\xi_4 \\ \bm{k}_c, \bm{\tilde{k}}_c, \bm{q}}}W^{\xi_1\xi_2\xi_3\xi_4}_{el, \bm{q}}c'^{\dagger}_{\xi_1, \bm{\tilde{k}}_c+\bm{q}}v^{\dagger}_{\xi_2, \bm{k}_c-\bm{q}}c_{\xi_3, \bm{k}_c}c_{\xi_4, \bm{\tilde{k}}_c}+\mathrm{h.c.} \ , 
\end{equation}
where $\xi_1...\xi_4$ are electronic valley indices and $\bm{k}_c, \bm{\tilde{k}}_c$ and $\bm{q}$ are momenta and $W^{\xi_1\xi_2\xi_3\xi_4}_{el,\bm{q}}$ is the electronic Auger matrix element  determining the strength of the recombination process. For simplicity, we omit the dependence on spin but note that it plays a similar role as the electronic valley index in the following calculation. Here, $c^{(\dagger)}$ and $v^{(\dagger)}$ describe the creation or annihilation operators for electrons and holes in the upper spin-split valence bands $v$ and the first spin-allowed conduction bands $c$, respectively. However, $c'^{(\dagger)}$ describes a creation or annihilation of an electron in a higher conduction band $c'$.  Note that we exclude \emph{hole-mediated} Auger scattering processes in this work, which incorporates transitions of holes between different valence bands. These processes are assumed to be inefficient due to the large energetic separations between valence bands in the considered monolayer.

We now consider the Auger contributions to the Heisenberg equation of motion for the microscopic interband polarisation  $ P^{\dagger \xi_e \xi_h}_{\bm{p+Q}, \bm{p}}=c^{\dagger}_{\xi_e, \bm{p+Q}}v_{\xi_h, \bm{p}}$ yielding
\begin{gather}
\begin{aligned}
     &i\hbar \partial_t P^{\dagger \xi_e \xi_h}_{\bm{p+Q}, \bm{p}}= [ P^{\dagger \xi_e \xi_h}_{\bm{p+Q}, \bm{p}}, H_{\mathrm{Aug}}]=\sum_{\xi_1...\xi_4, \bm{k}_c, \bm{\tilde{k}}_c, \bm{q}}W^{\xi_1\xi_2\xi_3\xi_4}_{el, \bm{q}}\bigg( c'^{\dagger}_{\xi_1, \bm{\tilde{k}}_c+\bm{q}}c_{\xi_4, \bm{\tilde{k}}_c}c^{\dagger}_{\xi_e, \bm{p+Q}}c_{\xi_3, \bm{{k}}_c}\delta^{\bm{p}, \bm{k}_c-\bm{q}}_{\xi_h, \xi_2}-\\
    &  -c'^{\dagger}_{\xi_1, \bm{\tilde{k}}_c+\bm{q}}v_{\xi_h, \bm{p}}v^{\dagger}_{\xi_2, \bm{k}_c-\bm{q}}c_{\xi_3, \bm{{k}}_c}\delta^{\bm{\tilde{k}}_c, \bm{p+Q}}_{\xi_4, \xi_e}
     -c'^{\dagger}_{\xi_1, \bm{\tilde{k}}_c+\bm{q}}c_{\xi_4, \bm{\tilde{k}}_c}\delta^{\bm{k}_c, \bm{p+Q}}_{\xi_3, \xi_e}\delta^{\bm{{k}}_c-\bm{q}, \bm{p}}_{\xi_2, \xi_h} +c'^{\dagger}_{\xi_1, \bm{\tilde{k}}_c+\bm{q}}v_{\xi_h, \bm{p}}v^{\dagger}_{\xi_2, \bm{k}_c-\bm{q}}c_{\xi_4, \bm{{\tilde{k}}}_c}\delta^{\bm{{k}}_c, \bm{p+Q}}_{\xi_3, \xi_e}
    \bigg) \ . 
\end{aligned} 
\end{gather} 
Next, we expand the electronic operators in terms of polarisations $P^{\dagger}=c^{\dagger}v$, $P'^{\dagger}=c'^{\dagger}v$. Here, we use the unit operator method introduced by Ivanov and Haug \cite{ivanov1993self} 
to transfer electronic operators to electron-hole pairs,
such that $c^{\dagger}c\sim P^{\dagger}P$ (neglecting higher order terms), with details provided in \cite{katsch2018theory, erkensten2021exciton}. This yields the following equation of motion
\begin{gather}
\begin{aligned}
    i\hbar \partial_t P^{\dagger \xi_e \xi_h}_{\bm{p+Q}, \bm{p}}&=\sum_{\xi_1...\xi_3, \bm{k}, \bm{q}}\bigg( W^{\xi_1\xi_h \xi_3\xi_e}_{el, \bm{q}}P'^{\dagger \xi_1 \xi_2}_{\bm{p+Q+q}, \bm{k}}P^{\xi_3 \xi_2}_{\bm{p+q}, \bm{k}}+W^{\xi_1\xi_2 \xi_e\xi_3}_{el, \bm{q}}P'^{\dagger \xi_1 \xi_h}_{\bm{k}+\bm{q}, \bm{p}}P^{\xi_3 \xi_2}_{\bm{k}, \bm{p+Q-q}}\\
    \label{heisen}
    & \ \ \ \ -W^{\xi_1\xi_2 \xi_3\xi_e}_{el, \bm{q}}P'^{\dagger \xi_1 \xi_h}_{\bm{p+Q+q}, \bm{p}}P^{\xi_3 \xi_2}_{\bm{k}, \bm{k-q}}-W^{\xi_1\xi_h \xi_e\xi_3}_{el, \bm{q}}\sum_{\bm{k'}}P'^{\dagger \xi_1 \xi_2}_{\bm{k'}+\bm{q}, \bm{k}}P^{\xi_3 \xi_2}_{\bm{k'}, \bm{k}}\delta_{\bm{q}, \bm{Q}}\bigg) \ . 
    \end{aligned}
    \end{gather}
Now, we express the polarisations in the excitonic basis yielding
\begin{equation}
    P^{\dagger \xi_1 \xi_2}_{\bm{k}_1, \bm{k}_2}=\sum_{n}X^{\dagger \xi_1 \xi_2}_{n, \bm{k}_1-\bm{k}_2}\varphi^{* \xi_1 \xi_2}_{n, \beta^{\xi_1\xi_2}\bm{k}_1+\alpha^{\xi_1\xi_2}\bm{k}_2} \ ,  P'^{\dagger \xi_1 \xi_2}_{\bm{k}_1, \bm{k}_2}=\sum_{n}Y^{\dagger \xi_1 \xi_2}_{n, \bm{k}_1-\bm{k}_2}\phi^{* \xi_1 \xi_2}_{n, \beta^{\xi_1\xi_2}\bm{k}_1+\alpha^{\xi_1\xi_2}\bm{k}_2}
    \label{excbasis}
\end{equation}
Note that we introduced different excitonic operators ($X^{(\dagger)}$ and $Y^{(\dagger)}$) and wave functions ($\varphi$ and $\phi$) for the ($c$,$v$) and ($c'$,$v$) excitons, respectively. In the following we will refer to ($c$,$v$) as the A exciton and ($c'$,$v$) as the HX (higher-lying exciton). Then, we can immediately rewrite the left-hand side in \eqref{heisen} 
as $ i\hbar \partial_t P^{\dagger \xi_e \xi_h}_{\bm{p+Q}, \bm{p}}=i\hbar \partial_t \sum_{n}X^{\dagger \xi_e \xi_h}_{n, Q}\varphi^{* \xi_e \xi_h}_{n, \beta^{\xi_e\xi_h}\bm{Q}+\bm{p}}$. To simplify notation, we introduce the compound index $\mu=(\xi, n)$. Moreover, we multiply the entire equation of motion written in exciton basis by $\varphi_{\sigma, \bm{q}_2}$ and sum over $\bm{q}_2=\beta^{\mu}\bm{Q}+\bm{p}$. Then the left-hand side converts to $i\hbar \partial_t \sum_{\mu}X^{\dagger}_{\mu, \bm{Q}}\varphi^{*}_{\mu, \bm{q}_2}\rightarrow i\hbar \partial_t X^{\dagger}_{\sigma, \bm{Q}}$ due to the orthogonality between excitonic wave functions. By  substituting \eqref{excbasis} in the right-hand side of \eqref{heisen}, multiplying by $\varphi_{\sigma, \bm{q}_2}$ and summing over $\bm{q}_2$, we find the equation of motion in excitonic basis
\begin{align}
    i\hbar \partial_t X^{\dagger}_{\sigma, \bm{Q}}=\sum_{\mu, \nu, \bm{Q'}}(E^{\mu\nu\sigma}_{\bm{Q}, \bm{Q}'}-D^{\mu\nu\sigma}_{\bm{Q}, \bm{Q}'})Y^{\dagger}_{\mu, \bm{Q}'+\bm{Q}}X_{\nu, \bm{Q}'} \ , 
    \label{excbas}
\end{align}
with the abbreviations
\begin{eqnarray*}
    E^{\mu\nu\sigma}_{\bm{Q}, \bm{Q}'}&\equiv&\sum_{\bm{q}, \bm{q}'} \bigg(W^{\mu_e\nu_h\sigma_e\nu_e}_{el, \bm{q-q'}}\phi^*_{\mu, \beta^{\mu\nu}\bm{Q}+\bm{q}-\alpha^{\mu}\bm{Q}
    }\varphi_{\nu, \bm{q}-\beta^{\nu}\bm{Q}-\alpha^{\sigma}\bm{Q}'}\varphi_{\sigma, \bm{q}'}\delta_{\mu_h, \sigma_h} + W^{\mu_e\sigma_h\nu_e\sigma_e}_{el, \bm{q-q'}}\phi^*_{\mu, \beta^{\mu\sigma}\bm{Q'}+\bm{q}-\alpha^{\mu}\bm{Q'}
    }\varphi_{\sigma, \bm{q}-\beta^{\sigma}\bm{Q'}-\alpha^{\nu}\bm{Q}}\varphi_{\nu, \bm{q}'}\delta_{\mu_h, \nu_h}\bigg) \ , \\
    D^{\mu\nu\sigma}_{\bm{Q}, \bm{Q}'}&\equiv& W^{\mu_e\nu_h\sigma_e\nu_e}_{el, \bm{Q}'}\sum_{\bm{q}, \bm{q}'}\phi^*_{\mu, \beta^{\mu\sigma}\bm{Q'}+\beta^{\mu}\bm{Q}+\bm{q}}\varphi_{\sigma, \bm{q}}\varphi_{\nu, \bm{q'}}\delta_{\mu_h, \sigma_h}+W^{\mu_e\sigma_h\nu_e\sigma_e}_{el, \bm{Q}}\sum_{\bm{q}, \bm{q}'}\phi^*_{\mu, \beta^{\mu\nu}\bm{Q}+\beta^{\mu}\bm{Q'}+\bm{q}}\varphi_{\nu, \bm{q}}\varphi_{\sigma, \bm{q'}}\delta_{\mu_h, \nu_h}
    \label{eqw}
\end{eqnarray*}
Here, $D$ and $E$ constitute direct and exchange interaction terms, respectively. Note that the matrix elements are symmetric with respect to the exchanges of initial A exciton states $\nu\leftrightarrow \sigma$ and momenta $\bm{Q}\leftrightarrow \bm{Q'}$.

Now, use directly  an excitonic Hamiltonian by performing the following ansatz:
\begin{equation}
    H_{x-x}=\frac{1}{2}\sum_{\mu, \nu, \sigma, \bm{Q}, \bm{Q}'}W^{\mu\nu\sigma}_{\bm{Q}, \bm{Q}'}Y^{\dagger}_{\mu, \bm{Q+Q'}}X_{\nu, \bm{Q'}}X_{\sigma, \bm{Q}} +\mathrm{h.c.}\ , 
    \label{ansatz}
\end{equation}
This way, we are able to read off the excitonic Auger matrix element $W^{\mu\nu\sigma}_{\bm{Q}, \bm{Q}'}$ from the corresponding Heisenberg equation of motion for the excitonic polarisation
\begin{equation}
    i\hbar\partial_t  X^{\dagger}_{\sigma, \bm{Q}}=[X^{\dagger}_{\sigma, \bm{Q}}, H_{x-x}]=-\sum_{\mu,\nu, \bm{Q'}}W^{\mu\nu\sigma}_{\bm{Q}, \bm{Q}'}Y^{\dagger}_{\mu, \bm{Q'+Q}}X_{\nu, \bm{Q}'} \ . 
    \label{excitonc}
\end{equation}
Note that here we have treated the operators $X^{(\dagger)}$ and $Y^{(\dagger)}$ as commuting bosonic operators. Still, note that the fermionic character of the excitons is retained and incorporated into the excitonic Auger matrix element $W$.
By comparison with \eqref{excbas} we immediately find the expression for the excitonic Auger matrix elements
\begin{equation}
    W^{\mu\nu\sigma}_{\bm{Q}, \bm{Q}'}=(D^{\mu\nu\sigma}_{\bm{Q}, \bm{Q}'}-E^{\mu\nu\sigma}_{\bm{Q}, \bm{Q}'}) \ , 
    \label{excmat}
\end{equation}
with the direct and exchange matrix elements being provided in \eqref{eqw}. Next, we note that the magnitude of the center-of-mass momenta $\bm{Q}, \bm{Q'}$ are of the same order as the thermal wave vector, $Q_T=\frac{2M k_B T}{\hbar^2}$ and that the relative exciton momentum $|\bm{q}|\sim a_B^{-1}$, $a_B$ being the exciton Bohr radius \cite{glazovprx}. In general, it holds that $Q_T<<a_B^{-1}$ and we may therefore neglect the center-of-mass momentum dependence entering the excitonic wave functions inside $D$ and $E$. Within this approximation we obtain the following simplified expressions for the direct and exchange matrix elements
\begin{gather}
\begin{aligned}
    E^{\mu\nu\sigma}_{\bm{Q}, \bm{Q}'}&\approx E^{\mu\nu\sigma}_{0,0}=\sum_{\bm{q}, \bm{q}'}(W^{\mu_e\nu_h\sigma_e\nu_e}_{el, \bm{q-q'}}\phi^*_{\mu, \bm{q}}\varphi_{\nu, \bm{q}}\varphi_{\sigma, \bm{q}'
    }\delta_{\mu_h, \sigma_h}+W^{\mu_e\sigma_h\nu_e\sigma_e}_{el, \bm{q-q'}}\phi^*_{\mu, \bm{q}}\varphi_{\sigma, \bm{q}}\varphi_{\nu, \bm{q}'
    }\delta_{\mu_h, \nu_h}) \\ 
       D^{\mu\nu\sigma}_{\bm{Q}, \bm{Q}'}&\approx W^{\mu_e\nu_h\sigma_e\nu_e}_{el, \bm{Q}'}F_{\mu\sigma}\tilde{\varphi}_{\nu, \bm{r}=0}\delta_{\mu_h, \sigma_h} +W^{\mu_e\sigma_h\nu_e\sigma_e}_{el, \bm{Q}}F_{\mu\nu}\tilde{\varphi}_{\sigma, \bm{r}=0}\delta_{\mu_h, \nu_h} \ , 
       \label{matelements}
\end{aligned}
\end{gather} 
with the form factor $F_{\mu\nu}=\sum_{\bm{q}}\phi^*_{\mu, \bm{q}}\varphi_{\nu, \bm{q}}$ and with $\sum_{\bm{q}}\varphi_{\nu, \bm{q}}=\tilde{\varphi}_{\nu, \bm{r}=0}$, where $\tilde{\varphi}_{\nu, \bm{r}}$ is the real-space representation of the A exciton wave function. 
In the main text we have introduced $G^{\mu\nu\sigma}=\sum_{\bm{q,q'}} G^{\mu\nu\sigma}_{\bm{q}, \bm{q'}}$ and $G^{\mu\nu\sigma}_{\bm{q}, \bm{q'}}=\phi^*_{\mu, \bm{q}}\varphi_{\nu, \bm{q}}\varphi_{\sigma, \bm{q'}}$ in the definition of the matrix elements to simplify notation. Having microscopic access to the excitonic Auger matrix element, we can now make use of the simpler Hamiltonian in \eqref{ansatz} when evaluating the exciton-exciton annihilation rate. 

\begin{figure}[t!]
        \centering
        \includegraphics[scale=0.3]{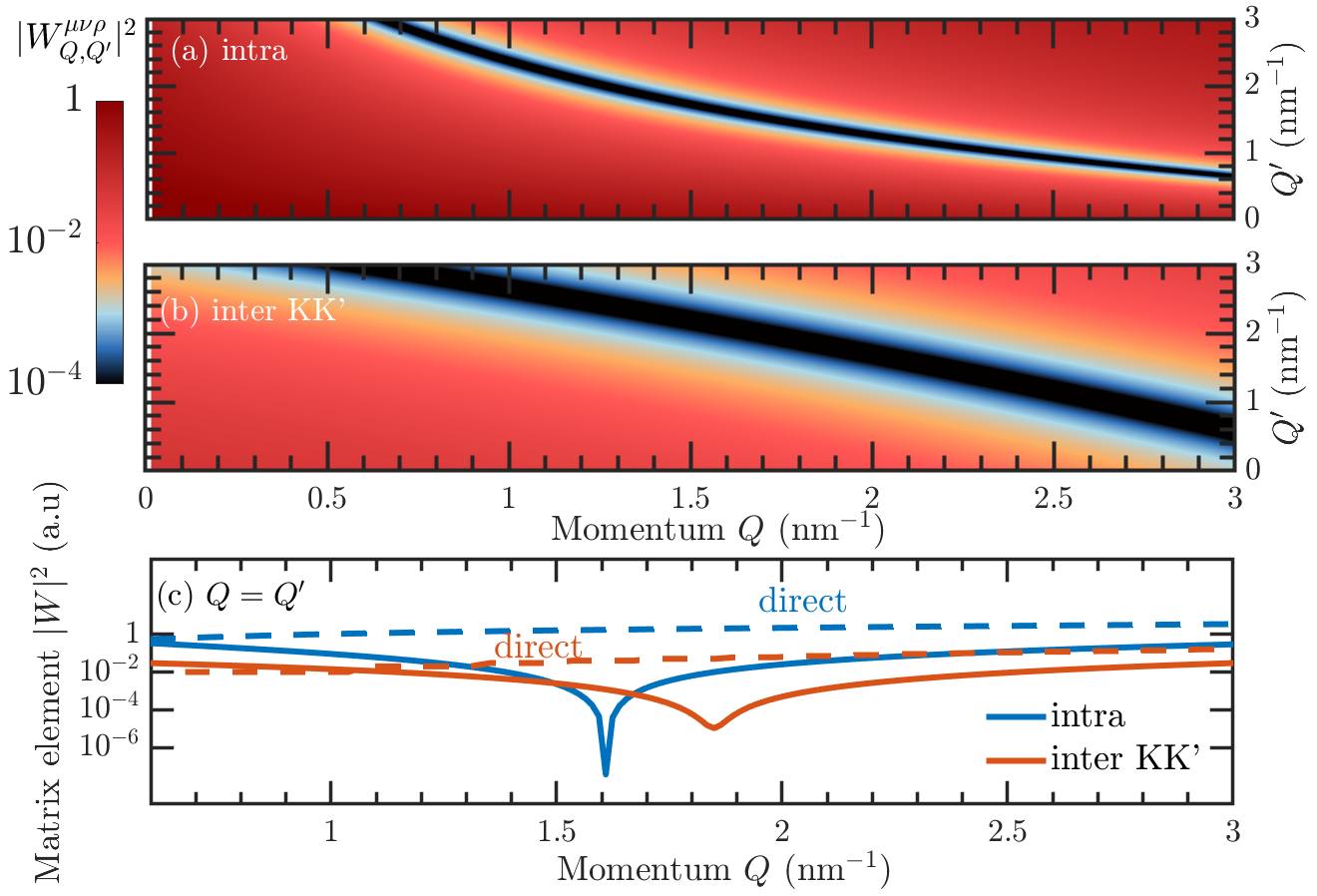}
        \caption{Momentum-dependent \textbf{(a)} intravalley and \textbf{(b)} intervalley Auger matrix element. \textbf{(c)} Separation of direct and exchange contributions in the Auger matrix elements. The dashed lines include only the direct term, while the solid lines include the full interaction.}
        \label{figur}
    \end{figure}

Finally, we define the \emph{electronic} Auger matrix element $W_{el, \bm{q}}$ entering \eqref{matelements}. Depending on the size of the momentum transfer $\bm{q}$, different matrix elements are obtained. For scattering processes with small momentum transfer $\bm{q}$, i.e. $\xi_{\mu}$=KK, $\xi_{\nu}, \xi_{\rho}$=KK$^{(')}$ we make use of $\bm{k}\cdot\bm{p}$ theory \cite{kane, luttinger} to find
\begin{equation}
    W^{\mu_e\nu_h\rho_e\nu_e}_{el, \bm{q}}=V_{\bm{q}}\bm{q}^2\gamma^{\rho_e-\nu_h}_{c-v}\gamma^{\nu_e-\mu_e}_{c-c'} \delta_{\rho_e, \nu_h}^{\mu_e, \nu_e} \ \ \ (\mathrm{intravalley})\ .     \label{intravalley}
\end{equation}
The matrix element is determined by the screened Coulomb interaction $V_{\bm{q}}$ and the electronic dipole matrix element $\gamma^{\eta-\eta'}_{\lambda-\lambda'}=\langle  \lambda, \eta |r|\lambda', \eta'\rangle$ describing transitions between different bands $\lambda, \lambda'=v,c,c'$ and valleys $\eta,\eta'=$ K, K'. For scattering processes with large momentum transfer $\tilde{\bm{q}}$ on the order of the reciprocal lattice vector, we decompose $\tilde{\bm{q}}=\bm{q}+\Delta_{\eta, \eta'}$, where $\bm{q}$ is small and $\Delta_{\eta, \eta'}$ denotes the momentum-difference between different high symmetry points $\eta\neq \eta'$ (i.e. $|\Delta_{\mathrm{K}, \mathrm{\Lambda}}|=\frac{2\pi}{3a_0}$ or $|\Delta_{\mathrm{K}, \mathrm{K'}}|=\frac{4\pi}{3a_0}$, $a_0$ being the lattice constant) and find
\begin{equation}
      W^{\mu_e\nu_h\rho_e\nu_e}_{el, \bm{q}}=V_{\bm{q}+\bm{\Delta}_{\nu_h, \rho_e}}S^{\rho_e-\nu_h}_{c-v}(\bm{q})S^{\nu_e-\mu_e}_{c-c'}(-\bm{q}) \ \ \ (\mathrm{intervalley}), 
      \label{intervalley}
\end{equation}
with $S^{\eta-\eta'}_{\lambda-\lambda'}(\bm{q})= (\tilde{\gamma}^{\eta-\eta'}_{0, \lambda-\lambda'}+i\bm{q} \tilde{{\gamma}}^{\eta-\eta'}_{1, \lambda-\lambda'})$, where  $\tilde{{\gamma}}_{0,\lambda-\lambda'}^{\eta-\eta'}=\langle \lambda, \eta|\lambda', \eta'\rangle$ and 
$\tilde{\bm{\gamma}}_{1,\lambda-\lambda'}^{\eta- \eta'}= \langle \lambda, \eta| \mathrm{e}^{i{\Delta}_{\eta, \eta'}r}r|\lambda', \eta'\rangle$, $\eta, \eta'=\mathrm{K}, \mathrm{K}', \mathrm{\Lambda}, \mathrm{\Gamma}$. Note that \eqref{intervalley} reduces to \eqref{intravalley} for $\eta=\eta'$. The band- and valley-specific $\gamma$-parameters are extracted from $GW$ (DFT) calculations described in detail in section III. Due to the large momentum transfer involved in the intervalley processes we do not employ the conventional Keldysh screening \cite{keldysh, rytova2018screened, cudazzo2011dielectric} entering the screened Coulomb potential, as it is seen to diverge in the short wave-length limit. Instead we make use of a non-linear screening function  \cite{thygesendft1, katsch2020exciton} in good agreement with ab-initio calculations for all momenta \cite{louiedft,thygesendft2}. 

In Fig. \ref{figur} (a)-(b) we illustrate the excitonic Auger matrix elements $W_{\bm{Q}, \bm{Q}'}$ defined in Eq. \eqref{excmat} for intra- and intervalley processes. We find that due to the large momentum transfer, the intervalley ($\xi_{\nu}=\xi_{\rho}=$KK', $\xi_{\mu}=$K$\Gamma$) matrix element is up to two orders of magnitude smaller than the corresponding intravalley ($\xi_{\nu}=\xi_{\rho}=\xi_{\mu}$=KK) matrix element. For large momentum transfer the electronic matrix elements are close to constant in momentum, resulting in a large cancellation of direct and exchange terms (cf. Eq. \eqref{matelements}), which explains the large dark-blue area in Fig.\ref{figur}(b). The direct and exchange contributions are further separated in Fig.\ref{figur} (c), where a cut of the matrix elements at $\bm{Q}=\bm{Q}'$ is shown. Considering only the direct contribution (dashed) we note a small monotonous increase with momentum. Including also  the exchange interaction  (solid), we observe a pronounced suppression of the matrix elements further reflecting the cancellation of the direct and the exchange term.

\subsection{Modeling of Auger coefficients}
In this section we provide a detailed derivation of the exciton-exciton annihilation rate expressed in the Auger coefficient $R_A$. We start from the  density of A excitons $n_{x}=\sum_{\nu, \bm{Q}}N^{\nu}_{\mathrm{A}, \bm{Q}}$, where the exciton occupation $N^{\nu}_{\mathrm{A}, \bm{Q}}=\langle X^{\dagger}_{\nu, \bm{Q}}X_{\nu, \bm{Q}}\rangle$ is estimated by a thermalized Boltzmann distribution. Here, $\bm{Q}$ is the center-of-mass momentum and $\nu$ is a compound index incorporating excitonic spin, valley (incl. valley degeneracy) and principal quantum number (which we restrict to be $n=1s$). Now, the goal is to derive an equation of motion for the density, which is non-linear in $n_{\mathrm{A}, x}$. By commuting the excitonic Auger Hamiltonian $H_{x-x}$ in \eqref{ansatz} with the occupation $N^{\nu}_{\mathrm{A}, \bm{Q}}$ we obtain the following Heisenberg equation of motion
\begin{equation}
   \partial_t N^{\nu}_{\mathrm{A}, \bm{Q}}= -\frac{2}{\hbar}\sum_{\substack{\mu, \rho, \bm{Q}'}}\mathrm{Im}\bigg( W^{\mu\nu\rho}_{\bm{Q}, \bm{Q}'}\langle Y^{\dagger}_{\mu, \bm{Q+Q'}}X_{\nu, \bm{Q}}X_{\rho, \bm{Q}'}\rangle \bigg) \ .
   \label{commut}
\end{equation}
 Then, we write $\langle Y^{\dagger}_{\mu, \bm{Q+Q'}}X_{\nu, \bm{Q}}X_{\rho,\bm{Q}'}\rangle=\mathrm{HF}+\langle Y^{\dagger}_{\mu, \bm{Q+Q'}}X_{\nu, \bm{Q}}X_{\rho, \bm{Q}'}\rangle^c$. We go beyond the Hartree-Fock approximation (HF) \cite{kira2006many, haug2009quantum} and find an equation of motion for the correlated part $C^{\mu\nu\rho}_{\bm{Q}, \bm{Q'}}\equiv\langle Y^{\dagger}_{\mu, \bm{Q+Q'}}X_{\nu, \bm{Q}}X_{\rho, \bm{Q}'}\rangle^c$ reading
\begin{equation}
    \partial_t C^{\mu\nu\rho}_{\bm{Q}, \bm{Q'}}\approx \frac{i}{\hbar}(\Delta \epsilon) C^{\mu\nu\rho}_{\bm{Q}, \bm{Q'}}+\frac{i}{\hbar}(W^{\mu\nu\rho}_{\bm{Q}, \bm{Q}'})^*N^{\nu}_{\mathrm{A}, \bm{Q}}N^{\rho}_{\mathrm{A}, \bm{Q}'} \ .
    \label{correl}
\end{equation}
Here, we have introduced the energy difference  $\Delta\epsilon=\Delta+\epsilon^{\mu}_{\mathrm{HX}, \bm{Q+Q'}}-\epsilon^{\nu}_{\mathrm{A}, \bm{Q}}-\epsilon^{\rho}_{\mathrm{A}, \bm{Q}'}$, with the \emph{detuning} $\Delta\equiv E_{\mathrm{HX}}-2E_{\mathrm{A}}+\varepsilon^{\mu}_{\mathrm{HX}}-\varepsilon^{\mathrm{KK},1s}_{\mathrm{HX}}+2\varepsilon^{\mathrm{KK}, 1s}_{\mathrm{A}}-\varepsilon^{\nu}_{\mathrm{A}}-\varepsilon^{\rho}_{\mathrm{A}}$, where the A and HX resonances can be obtained from recent PL measurements \cite{lin2020bright} and the eigenenergies $\varepsilon^{\nu}_{\mathrm{A}}$ and $\varepsilon^{\mu}_{\mathrm{HX}}$ are extracted from the Wannier equation (cf. Section C). Note that $\Delta=E_{\mathrm{HX}}-2E_{\mathrm{A}}$ for intravalley Auger scattering. We make use of the parabolic approximation for the A and HX excitons, i.e. $\epsilon^{\mu}_{\bm{Q}}=\frac{\hbar^2 \bm{Q}^2}{2M^{\xi_{\mu}}}$, $M$ being the total exciton mass. 

We also performed the mean-field and random phase approximation according to $\langle X^{\dagger}_{1}X^{\dagger}_{1}X_{3}X_{4}\rangle \approx \langle X^{\dagger}_1 X_3\rangle\langle X^{\dagger}_2 X_4\rangle\delta^{1,3}_{2,4}+\langle X^{\dagger}_1 X_4\rangle\langle X^{\dagger}_2 X_3\rangle\delta^{1,4}_{2,3}$. We assumed the HX occupation $N^{\mu}_{\mathrm{HX}, \bm{Q}}=\langle Y^{\dagger}_{\mu, \bm{Q}}Y_{\mu, \bm{Q}}\rangle$ to be negligible and only kept terms non-linear in the A exciton density. Next, we solve the equation of motion \eqref{correl} within a Markov approximation \cite{haug2009quantum} yielding $C^{\mu\nu\rho}_{\bm{Q}, \bm{Q}'}(t)\approx (W^{\mu\nu\rho}_{\bm{Q}, \bm{Q}'})^*N^{\nu}_{\mathrm{A}, \bm{Q}}N^{\rho}_{\mathrm{A}, \bm{Q}'}(i\pi\delta(\Delta\epsilon)+\mathcal{P}(1/\Delta\epsilon))$ with the delta function $\delta(\Delta \epsilon)$ reflecting energy conservation and the real-valued principal-value integral $\mathcal{P}(1/\Delta\epsilon)$ leading to many-body energy renormalizations. The latter plays no role in scattering processes and cancels out when plugging in $C^{\mu\nu\rho}_{\bm{Q}, \bm{Q'}}(t)$ into \eqref{commut}. 
Summing \eqref{commut} over $\bm{Q}$ and $\nu$ leads to an equation of motion for the density $n_{\mathrm{A,x}}$ in the incoherent limit reading $\partial_t n_{\mathrm{A}, x}=-R_A n_{\mathrm{A}, x}^2 $ with the Auger coefficient 
\begin{equation} R_A=\frac{2\pi}{\hbar}\sum_{\substack{\mu\nu\rho\\ \bm{Q}\bm{Q}'}}|W^{\mu\nu\rho}_{\bm{Q}, \bm{Q}'}|^2\bar{N}^{\nu}_{\mathrm{A}, \bm{Q}}\bar{N}^{ \rho}_{\mathrm{A}, \bm{Q}'}\delta(\Delta\epsilon) \ ,  
\end{equation}
with $\bar{N}=N/n_x$. Similarly, it is possible to find an equation of motion for the HX density $n_{\mathrm{HX}, x}=\sum_{\mu, \bm{Q}}\langle Y^{\dagger}_{\mu, \bm{Q}}Y_{\mu, \bm{Q}}\rangle$ according to $\partial_t n_{\mathrm{HX}, x}=\frac{R_A}{2} n_{\mathrm{A}, x}^2$ \cite{glazovprx}.

\subsection{Excitonic binding energies}
We microscopically calculate excitonic  binding energies and wave functions by solving the Wannier equation
\begin{equation}
    \frac{\hbar^2 k^2}{2m^{\xi_{\nu}}}\varphi_{\nu, \bm{k}}-\sum_{\bm{q}}W_{\bm{q}}\varphi_{\nu, \bm{k+q}}=\varepsilon^{\nu}\varphi_{\nu, \bm{k}} \ ,
\end{equation}
with $m^{\xi_{\nu}}=\frac{m^{\nu_e}m^{\nu_h}}{m^{\nu_e}+m^{\nu_h}}$ being the reduced mass in valley $\xi_{\nu}=$KK, KK', K$\mathrm{\Lambda}$, K$\mathrm{\Gamma}$. Here, $\varepsilon^{\nu}$ is the excitonic binding energy for the state $\nu=(\xi_{\nu}, n)$, $n=1s,2s...$ being the principal quantum number. Note that we include $n=1s \ (1s,2s,3s)$ A (HX) states in our calculations.
To obtain the energetic position of excitonic energies, $E_{\nu}$, we also include the single-particle separation $E^{0}_{\xi_{\nu}}$ between the top of the valence band at the K-point and the bottom of the first spin-allowed conduction bands at K, K' and $\mathrm{\Lambda}$, such that $E_{\nu}=E^{0}_{\xi_{\nu}}+\varepsilon^{\nu}$. For the higher-lying exciton (HX) we also include the single-particle
separation between the first conduction band ($c$) and the higher-lying conduction band ($c'$). However, when computing the detuning we eliminate the quasiparticle band gaps at the K-point by using $E_A=E_g+\varepsilon^{\mathrm{KK},1s}_{\mathrm{A}}$ and $E_{HX}=\tilde{E}_g+\varepsilon^{\mathrm{KK}, 1s}_{\mathrm{HX}}$ and adapting the A and HX resonances from recent up-converted PL measurements. 
\begin{table}[t!]
\begin{tabular}{|l|l|l|}
\hline
Exciton energies WSe$_2$ ($\epsilon_s=4.5$)  & $l=\mathrm{A}$ (meV) & $l=\mathrm{HX}$ (meV) \\ \hline
$\varepsilon_l(\mathrm{KK},1s)$         & -160        & -500         \\ \hline
$\Delta^l_{\mathrm{KK}, \mathrm{K\Lambda}}$  & -39         & -            \\ \hline
$\Delta^l_{\mathrm{KK}, \mathrm{KK'}}$       & -53         & 331          \\ \hline
$\Delta^l_{\mathrm{KK}, \mathrm{K\Gamma_1}}$ & -           & -6           \\ \hline
$\Delta^l_{\mathrm{KK}, \mathrm{K\Gamma_2}}$ & -           & 56           \\ \hline
\end{tabular}
\caption{1s exciton energies in hBN-encapsulated monolayer WSe$_2$ for A ($c,v$) and HX ($c',v$) excitons. Here, $\Delta^{l}_{\mathrm{KK}, \mu}$ denotes spectral exciton separations with the total (binding) energy given by $\varepsilon^{\mu}_l=\varepsilon_l(\mathrm{KK}, 1s)+\Delta^{l}_{\mathrm{KK}, \mu}$ (relative to the single-particle band gap). }
\label{tabell}
\end{table}
In Table \ref{tabell} we provide the 1s A and HX exciton energies. Note that conduction bands at the $\Gamma$-point are degenerate and have no preferred spin polarisation. However, the bands have different curvature (cf. Table \ref{masstable}) and therefore both bands are explicitly included in the calculations.
\newpage 
\section{First-principle parameters for microscopic theory}
The effective masses and \textbf{k}$\cdot$\textbf{p} matrix elements necessary for the microscopic theory were derived using a modern ab initio approach. The relevant quasiparticle energies were derived from one-shot G$_0$W$_0$ calculations as implemented in the YAMBO code~\cite{yambo}, using 2500 unoccupied bands in combination with the effective energy technique (EET)~\cite{eet} to account for higher-energy interband transitions. The band energies were extrapolated to infinite cutoff energy for the correlation contribution with the method we employed previously in Ref.\onlinecite{Gillen2018PRB} and calculations for cutoff energies of 250\,eV, 300\,eV and 350\,eV, for which the variation of band energies is linear with the inverse of the number of included reciprocal lattice vectors. The effect of spin-orbit interaction was explictly accounted for. A homogeneous grid of 9x9x1 k-points in the Brillouin zone was found sufficient to capture the dispersion of the electronic quasiparticles. The quasiparticle band gap was then corrected by applying a scissor shift to the conduction bands. For this step, we derived the band gap correction without explicit inclusion of spin-orbit interaction, using the "Nonuniform Neck Subsampling" method~\cite{daJornada2017PRB} as implemented in the BerkeleyGW package~\cite{BerkeleyGW}. We found that this method was sufficient to converge the electronic band gap to less than 0.05\,eV. The quasiparticle band structures were then obtained through Wannier interpolation using the Wannier90~\cite{wannier90} code. The input electronic energies and spinors and the groundstate density for the GW and Wannier calculations were computed with the Quantum Espresso code~\cite{qe}, optimized norm-conserving Vanderbilt pseudopotentials~\cite{oncv}, a cutoff energy of 90\,Ry, a 12x12x1 k-point grid (for the groundstate calculations) and the Perdew-Burke-Ernzerhof (PBE)~\cite{PBE} exchange-correlation functional. We fully optimized the atomic positions and cell parameters until the residual forces between atoms and the cell stress were smaller than 0.01\,eV/\AA~and 0.01\,GPa, respectively. Long-range non-covalent interactions were included through the semi-empirical DFT-D3 correction with Becke-Johnson damping~\cite{DFT-D3}. 
Using the obtained Wannier Hamiltonian, we interpolated the G$_0$W$_0$ quasiparticle energies on a dense regular grid with within a radius of 0.15\,1/\AA\space around the high symmetry points. We then derived the effective masses of the valence and conduction bands valleys of interest by fitting a parabolic function to the obtained quasiparticles energies of WSe$_2$. The resulting effective masses are shown in Table~\ref{masstable} and generally in good agreement with the previously reported masses using the G$_0$W$_0$ approximation~\cite{Rybkovskiy2017PRB} or hybrid functionals within the DFT framework~\cite{Kormanyos2015k}.

The electronic spinors of the valence and conduction bands, which are readily available from our DFT calculations, further allowed us to extract the form factors for interband transitions from an initial state with band index $m$ and quasi-crystal momentum $\mathbf{k}$ to a final state $n$,$\mathbf{k}$+$\mathbf{Q}$. Assuming that we can neglect the momentum dependence of the Bloch wavefunctions for a small perturbation in momentum $\mathbf{q}$, i.e. $\left| n,\mathbf{k}+\mathbf{Q}+\mathbf{q}\right\rangle\approx \left|n,\mathbf{k}+\mathbf{Q}\right\rangle$, we can further expand the form factor in $\mathbf{q}$, yielding the expression
\begin{align}
\Gamma_{n,m}^{\mathbf{k}+\mathbf{Q},\mathbf{k}}\left(\mathbf{q}\right)\nonumber\\
                 &=       \langle n, \mathbf{k}+\mathbf{Q}| e^{i(\mathbf{Q}+\mathbf{q})\cdot\mathbf{r}} |m, \mathbf{k}\rangle\nonumber\\
                 &\approx \langle n, \mathbf{k}+\mathbf{Q}| e^{i\mathbf{Q}\cdot\mathbf{r}}(1+i\mathbf{q}\cdot\mathbf{r})|m, \mathbf{k}\rangle\nonumber\\
                 &\approx \langle n,  \mathbf{k}+\mathbf{Q}| e^{i\mathbf{Q}\cdot\mathbf{r}}|m,\mathbf{k}\rangle + i\mathbf{q}\cdot\langle n, \mathbf{k}+\mathbf{Q}| e^{i\mathbf{Q}\cdot\mathbf{r}}\mathbf{r}|m, \mathbf{k}\rangle\nonumber\\
                 &\approx \tilde{\gamma}_{0,nm}^{\mathbf{k}+\mathbf{Q},\mathbf{k}} + i\mathbf{q}\cdot \mathbf{\tilde{\gamma}}_{1,nm}^{\mathbf{k}+\mathbf{Q}, \mathbf{k}}\nonumber
                 \end{align}
For sufficiently small momentum $\mathbf{q}$, the $\mathbf{\tilde{\gamma}}$ parameters should be nearly isotropic. As an additional simplification, we hence decided to use the average of the x- and y-components of $\mathbf{\tilde{\gamma}}$ instead of the vectorial form:
\begin{equation}
\Gamma_{n,m}^{\mathbf{k}+\mathbf{Q},\mathbf{k}}\left(\mathbf{q}\right) \approx \tilde{\gamma}_{0,nm}^{\mathbf{k}+\mathbf{Q},\mathbf{k}} + i\left|\mathbf{q}\right|\cdot \tilde{\gamma}_{1,nm, ave}^{\mathbf{k}+\mathbf{Q},\mathbf{k}}\nonumber
\end{equation}

Using the completeness relation $I=\sum_l\left|l, \mathbf{k}\left\rangle\right\langle l, \mathbf{k}\right|$, we recast $\mathbf{\tilde{\gamma}}_{1,nm}^{\mathbf{k}+\mathbf{Q},\mathbf{k}}$ into the form
\begin{align}
\mathbf{\tilde{\gamma}}_{1,nm}^{\mathbf{k}+\mathbf{Q},\mathbf{k}} &= \left\langle n, \mathbf{k}+\mathbf{Q}\left|e^{i\mathbf{Q}\cdot\mathbf{r}}\mathbf{r}\right|m, \mathbf{k}\right\rangle\nonumber\\
   &= \left\langle n, \mathbf{k}+\mathbf{Q}\right| e^{i\mathbf{Q}\cdot\mathbf{r}}\left(\sum_l\left|l, \mathbf{k}\left\rangle\right\langle l, \mathbf{k}\right|\right)\mathbf{r}\left|m, \mathbf{k}\right\rangle\nonumber\\
   &= \sum_l\left\langle n, \mathbf{k}+\mathbf{Q}\left| e^{i\mathbf{Q}\cdot\mathbf{r}}\left|l, \mathbf{k}\right\rangle\left\langle l, \mathbf{k}\right|\mathbf{r}\right|m, \mathbf{k}\right\rangle\nonumber\\
   &= \sum_l \tilde{\gamma}_{0,nl}^{\mathbf{k}+\mathbf{Q},\mathbf{k}}\left\langle l, \mathbf{k}\left|\mathbf{r}\right|m, \mathbf{k}\right\rangle,\nonumber
\end{align}
which only contains terms that are straight-forwardly derived from the quasiparticle bandstructure and the calculated DFT spinors. In practice, we truncate the sum over $l$ after the first 250 bands, which we found to yield sufficiently converged values. Following the typical procedure, the matrix elements of the position operator can be approximated by the matrix elements of the momentum operator $\mathbf{p}=-i\hbar\nabla$, with $\left\langle \eta\left|\mathbf{r}\right|\eta'\right\rangle\approx -i\frac{\mathbf{\hbar}}{m\left(\epsilon_{\zeta}-\epsilon_{\zeta'}\right)}\left\langle \zeta\left|\mathbf{p}\right|\zeta'\right\rangle$. The momentum operator matrix elements are easily calculated from the Fourier representation of the electronic spinors. 
For the calculation of the $\gamma_0$ parameters, we make use of the relation
\begin{equation}
A_{n\mathbf{k},m\mathbf{k}'}\left(\mathbf{\tilde{G}}\right)=\left\langle n,\mathbf{k}\left| e^{i\left(\mathbf{Q}+\mathbf{\tilde{G}}\right)\cdot\mathbf{r}} \right|m,\mathbf{k}'\right\rangle =\mathcal{F}\left[u_{n,\mathbf{k}}^*\left(\mathbf{r}\right)u_{m,\mathbf{k}'}\left(\mathbf{\mathbf{r}}\right)\right],
\end{equation}
where $u_{n,\mathbf{k}}$ are the lattice-periodic parts of the Bloch functions of band $n$ and crystal momentum $\mathbf{k}$. With this, $\tilde{\gamma}_{0,nl}^{\mathbf{k},\mathbf{k}}=A_{n\mathbf{k},l, \mathbf{k}'}\left(\mathbf{\tilde{G}}=\mathbf{0}\right)$. The resulting $\gamma$ parameters for monolayer WSe$_2$ are listed in Table~\ref{gammatable}.
\begin{table}[h!]
\begin{tabular}{|l|l|l|l|l|}
\hline
Eff. mass ($\times m_0$) & K             & K'    & $\mathrm{\Lambda}$ & $\mathrm{\Gamma}$ \\ \hline
$m^{\nu_e}_c$            & 0.31          & 0.4   & 0.6                & -                 \\ \hline
$m^{\mu_e}_{c'}$         & -0.52 (-0.46 \cite{lin2020bright}) & -0.46 & -                  & -0.84 (-1.74)     \\ \hline
$m^{\nu_h}_v$            & 0.4           & -     & -                  & -                 \\ \hline
\end{tabular}
\caption{Effective masses extracted from GW (DFT) calculations for the upper spin-split band $c$ and higher-lying $c'$ (also referred to as the $c+2$-band \cite{lin2020bright}) in monolayer WSe$_2$. We note a good agreement with previous ab-initio studies \cite{Kormanyos2015k}. }
\label{masstable}
\end{table}
\begin{table}[h!]
\begin{tabular}{|l|l|}
\hline
$\gamma$-parameters                                  & $|\gamma|$    \\ \hline
$\gamma^{\mathrm{K-K}}_{c-v}$                        & 3.2      \\ \hline
$\gamma^{\mathrm{K^{(')}-K^{(')}}}_{c-c'}$           & 0.58     \\ \hline
$\tilde{\gamma}^{\mathrm{\Lambda-K}}_{0, c-v}$       & 0.344         \\ \hline
$\tilde{\gamma}^{\mathrm{\Lambda-K}}_{1, c-v}$       & 0.558     \\ \hline
$\tilde{\gamma}^{\mathrm{\Lambda-\Gamma}}_{0, c-c'}$ & 0.244 (0.208) \\ \hline
$\tilde{\gamma}^{\mathrm{\Lambda-\Gamma}}_{1, c-c'}$ & 0.232 (0.198) \\ \hline
$\tilde{\gamma}^{\mathrm{K'-K}}_{0, c-v}$            & 0.252         \\ \hline
$\tilde{\gamma}^{\mathrm{K'-K}}_{1, c-v}$            & 0.95          \\ \hline
$\tilde{\gamma}^{\mathrm{K'-\Gamma}}_{0, c-c'}$      & 0.029 (0.034) \\ \hline
$\tilde{\gamma}^{\mathrm{K'-\Gamma}}_{1, c-c'}$      & 0.105 (0.124) \\ \hline
\end{tabular}
\caption{kp $\gamma$-parameters extracted from ab-initio calculations for different intra-and interband transitions in monolayer WSe$_2$. We distinguish intravalley and intervalley by the parameters $\gamma$ and $\tilde{\gamma}$ respectively. Note that we only provide the absolute magnitude of the parameters, since the parameters can be determined up to a phase. Moreover, note that $\tilde{\gamma}_1$ is formally a two-dimensional vector, but we find that it can be well approximated by the average of its $x$ and $y$ components when evaluating the electronic Auger matrix elements. For the transitions to the $\Gamma$-point we provide two numbers: one number for each spin-split band. Note that, due to time-reversal symmetry, bands at the $\Gamma$-point have no preferred spin polarisation and therefore we include both bands in the calculation of the Auger rates. All values are given in units of Å ($\mathrm{10}^{-10}$ m), except for the unitless $\gamma_0$.}
\label{gammatable}
\end{table}

\newpage